\documentclass[11pt,a4paper]{article}
 \usepackage{jheppub}
\usepackage{amsmath,amssymb}
\usepackage{epsfig}
\usepackage{dsfont}

\def\x'{\mathaccent 19 x}
\def\y'{\mathaccent 19 y}
\def\n'{\mathaccent 19 n}
\def\u'{\mathaccent 19 u}

\def\et'{\mathaccent 19 \eta}
\def\th'{\mathaccent 19 \theta}
\def\lam'{\mathaccent 19 \lambda}
\def\varet'{\mathaccent 19 \vartheta}
\def\rh'{\mathaccent 19 \rho}
\def\ph'{\mathaccent 19 \Phi}
\def\xb'{\mathaccent 19 {\bar{x}}}

\def\be{\begin{equation}}
\def\ee{\end{equation}}

\newcommand{\bea}{\begin{eqnarray}}
\newcommand{\eea}{\end{eqnarray}}

\author[a]{Fabrizio Pucci}
\affiliation[a]{Fakult\"{a}t f\"{u}r Physik, Universit\"{a}t Bielefeld,
Universit\"{a}tsstra{\ss}e 25, D-33615 Bielefeld, Germany}
\emailAdd{pucci@physik.uni-bielefeld.de}

\abstract{We study supersymmetric 't Hooft loop operators in ${\cal N}=4$ super Yang-Mills, generalizing the well-known circular 1/2 BPS
case and investigating their S-duality properties. We derive the BPS condition for a generic line operator describing pointlike
monopoles and discuss its solutions in some particular case. In particular, we present the explicit construction of the magnetic
counterpart of Zarembo and DGRT Wilson loops and provide the general dyonic configurations for an abelian gauge group.
The quantum definition of these supersymmetric 't Hooft loop operators is carefully discussed and we attempt some computations
to next-to-leading order in perturbation theory.
}
\subheader{\hspace{13cm}\small BI-TP 2012/34}

\title{More on 't Hooft loops in ${\cal N}$ = 4 SYM}
\begin{document}
\maketitle
\renewcommand{\thefootnote}{\arabic{footnote}}
\setcounter{footnote}{0}

\section{Introduction}

Four dimensional ${\cal N}=4$ supersymmetric Yang-Mills (SYM) theory still represents a fascinating playground to study dynamical properties of quantum fields at nonperturbative level. The rich mathematical structure and the large amount of supersymmetry constrain the theory enough to make it amenable to exact treatment. As an example, particular classes of loop operators, that are some of the most important observables in ${\cal N}=4$ SYM, have been exactly calculated in a series of nice papers \cite{Pestun:2007rz,Gomis:2011pf,TAKI} using localization techniques. Loop operators can be classified according to whether the running charges are electric or magnetic and describe heavy probe particles that move along a closed path in spacetime. Both classes are usually defined in any gauge theory respectively as order and disorder parameters and are useful to characterize the phases of the theory.\newline
The analysis of loop operators in supersymmetric theories has been a wide field of research over the last fifteen years, starting from the seminal papers on the ${\cal N}$=4 electric case  \cite{Rey:1998ik,Maldacena:1998im}. In maximally supersymmetric SYM theory these operators become extremely useful for two further reasons:  firstly their quantum expectation values provide interesting tools to check the AdS/CFT correspondence and indeed, for particular examples, exactly interpolating functions between weak and strong coupling regimes have been constructed. Furthermore they are also important in testing the action of $S$-duality, an exact quantum symmetry under which ${\cal N}$= 4 SYM is believed to be invariant. Since this duality exchange the role of the electric and the magnetic degrees of freedom it maps electric operators (Wilson loops) to their magnetic counterparts ('t Hooft loops) and relates their expectation values.\newline
The Wilson loop operator is usually defined as
\begin{equation}\langle W(C)\rangle = \frac{1}{{\rm dim}(R)} {\rm Tr}_R \left[{\cal P}\exp\left(i \oint_c A_{\mu} dx^{\mu}\right)\right] \label{10}
\end{equation}
\noindent
and essentially measures the response of the gauge field to an external quark-like source passing around a closed contour $C$\footnote{Above $R$ denotes the representations of the gauge group, where the quark-like external source transforms.}: in ordinary QCD loops in the fundamental representation are used to distinguish the different phases of the theory.
In ${\cal N}=4$ SYM the Wilson loop has been instead defined in \cite{Rey:1998ik, Maldacena:1998im} as

\begin{equation}\langle W(C)\rangle = \frac{1}{{\rm dim}(R)} {\rm Tr}_R\left[ {\cal P} \exp\left( \oint_C \left( i A_{\mu} \dot{x}^{\mu} + \phi^A \theta^A(s) |\dot{x}|\right) ds \right)\right]\label{MaldacenaWilson}\end{equation}
\noindent
and quite naturally the operator obtained couples not only with the gauge field but also with the six scalars of the theory. A necessary condition to preserve locally some amount of supersymmetry is that the couplings $\theta^A(s)$ ($A=1....6$) in
(\ref{MaldacenaWilson}) satisfy the constraint $\theta^A \theta^A = 1$ \cite{Zarembo:2002an}.
In order to have a global BPS object the following and more stringent condition has to be satisfied
\begin{equation} \left(\, i \Gamma_{\mu} \dot{x}_{\mu}(s) + \Gamma^A \theta^A(s) |\dot{x}|\, \right)\, \epsilon(s) = 0 \label{refffA}\end{equation}
namely eq. (\ref{refffA}) must admit a non trivial solution for

\begin{equation}
\epsilon(s)=\epsilon_0+x^\mu(s)\Gamma_\mu\epsilon_1,
\end{equation}
\noindent
that is a conformal Killing spinor on $\mathbb{R}^4$ with $\epsilon_0$ and  $\epsilon_1$ two sixteen-component Majorana-Weyl constant spinors. This requirement yields constraints either on the loop ($\dot{x}_{\mu}(s)$) or on the scalar couplings ($\theta^A(s)$)
or on both quantities.\newline
The most famous example of Wilson operator that satisfies the previous condition is the 1/2 BPS circular Wilson loop: it was
suspected for a long time that a matrix model computation should be capture all the information of this operator \cite{Er,Drukker:2000rr}.
The conjecture has been rigorously proven in \cite{Pestun:2007rz}, where it has been shown that the path-integral of theory defined on $S^4$ localizes on a finite dimensional space and reduces to a simple gaussian matrix model. The circular Wilson loop, due to its invariance properties, can be computed as an observable in this matrix model, leading to the expected result.\newline
A simple idea to solve eq. (\ref{refffA})  and obtain supersymmetric Wilson operators was proposed by Zarembo in \cite{Zarembo:2002an}: this construction is based on the additional requirement  that the position of the loop on the scalar $S^5$, defined by the functions $\theta^A(s)$, follows the tangent vector to the contour $C$. In this case one can show that for an arbitrary shape of the contour the loop preserves  1/16 of the original  Poincar\`e supersymmetry but if the curve lies in a  lower dimensional subspace of $\mathbb{R}^4$ we observe an enhancement of the supersymmetry.\newline
Another interesting proposal has been put forward by Drukker, Giombi, Ricci and Trancanelli  (DGRT) in \cite{Drukker:2007dw,Drukker:2007yx,Drukker:2007qr}.  There, the authors considered an new class of Wilson loops of  arbitrary shape defined on a space-time three sphere $S^3$:  for a generic curve these loops preserve two supercharges but they discussed special cases in which also 4 and 8 supercharges are conserved. Of particular interest are the loops restricted to $S^2$ because perturbative computations suggest
the equivalence with analogous observables in the purely bosonic two dimensional Yang-Mills theory on the two-sphere: there are different indications that this conjecture holds \cite{Pestun,PestunII,Nostro,Young,CorrelatoriI,CorrelatoriII} even if a complete proof of the conjecture is still missing (see \cite{Pestun} for more details). The above families  do not exhaust all the possible supersymmetric Wilson loops with only bosonic couplings. Recently a systematic classification of all possible solutions to the equation (\ref{refffA}) has been obtained  by Pestun and Dymarsky \cite{Dymarsky:2009si} and new kinds of BPS observables have been defined.\newline
In gauge theories it is also possible to introduce a completely different class of non-local observables, the disorder operators suggested by 't Hooft and commonly referred to as 't Hooft loops \cite{thooft}. They are defined by specifying the singularity that the field configurations have near a path $C$ on which the operators are supported and can be thought as the duals of the Wilson loops. Physically they inserts a probe point-like monopole whose world-line is the given loop $C$ and can be used to study the different phases of the theory as well. As shown in \cite{thooft}, while the expectation value of a Wilson loop operator in the confined phase satisfied the area law, the $VEV$ of the 't Hooft loop satisfies a perimeter law, and viceversa in the deconfined phase. The prescription to calculate the $VEV$ of these magnetic objects is to perform the path integral in the presence of the singularity along the path $C$, in the simple abelian case being the one associated to a Dirac monopole

\begin{equation}F_{0i}=0 \, \, \, \, \, \, \hspace{1cm} F_{ij} = \frac{m}{2} \epsilon_{ijk} \frac{x_i}{r^3},\end{equation}
\noindent
where $i,j$=1,2,3 and  $m \in \mathbb{Z}$ is the magnetic charge of the monopole. In the non-abelian case with arbitrary gauge group $G$ one embeds the construction by defining an homomorphism $\rho : U(1)\rightarrow G$ and requiring that the operator has along the loop a singularity that is the image under $\rho$ of the abelian one. More in detail, the configuration for a 't Hooft line operator is given by
\begin{equation}F_{0i}= i g^2 \frac{\theta}{16 \pi^2} B \frac{x_i}{r^3}\, \, \, \, \hspace{1cm} F_{ij} = \frac{B}{2} \epsilon_{ijk} \frac{x_i}{r^3}\label{nbps}\end{equation}
\noindent
where $B$ belongs to the Cartan subalgebra of $g$, the Lie algebra associated to the group $G$, and where we have also
taken into account that, in the presence of a non zero $\theta$ angle, an electric field is generated via the famous Witten effect \cite{dyn}.
Since the element $B$ has to obey to the Dirac quantization condition

\begin{equation}\mathrm{exp}(2 \pi i B)= \mathbf{1},\end{equation}
\noindent
one can show \cite{GNO} that it can be identified as a coweight vector (magnetic weight) and takes values in the coweight lattice $\Lambda_{cw}$ of $G$\footnote{Or equivalently in the weight lattice $\Lambda_{w}$ of the dual group $^L G$}. In ${\cal N}$ = 4 SYM these operators can be defined as well, but to obtain classical configurations that preserve a part of the supercharges we have to ensure the vanishing of the following supersymmetric variation

\begin{equation}\delta_{\epsilon} \Psi = \frac{1}{2}\Gamma^{MN} F_{MN}\, \epsilon(s) - 2 \phi^A \Gamma^A \epsilon_1 = 0\end{equation}
\noindent
for a non-trivial Killing spinor $\epsilon(s)$. Its solution implies that also the scalar fields have a singularity that goes like $1/r$ near the loop $C$: for example one can easily verify that the singular configurations associated to the supersymmetric version of the straight line operator (\ref{nbps}) are given by

\begin{equation}F_{0i}= i g^2 \frac{\theta}{16 \pi^2} B \frac{x_i}{r^3}\, \, \, \, \, \, \, F_{ij} = \frac{B}{2} \epsilon_{ijk} \frac{x_i}{r^3}\, \, \, \, \, \,  \phi^9 = g^2 \frac{B}{8 \pi} |\tau| \frac{1}{r}\label{conf},\end{equation}
\noindent
where only one of the six scalars of the theory is turned on and

\begin{equation}\tau = \frac{\theta}{2 \pi} + \frac{4 \pi i}{g^2}\label{tau}\end{equation}
\noindent
is the generalized coupling constant. A well known non-trivial BPS magnetic operator is the circular 't Hooft loops. It has been studied carefully first in \cite{Okuda}, where the definition of the quantum operator and the prescription to calculate its expectation value have been given. More recently in \cite{Gomis:2011pf} the calculation of its $VEV$ has been performed exactly with a localization procedure on $S^4$: compared to the circular Wilson Loop's computation there is a new and crucial contribution arising from the equator of $S^4$ where the loop is supported. Furthermore in \cite{Gomis:2009xg} the OPE analysis of the circular loop and the correlation functions with an arbitrary chiral primary operator have been studied.\newline
The knowledge of the exact expression for the $VEV$ of both Wilson and 't Hooft operators opens the possibility to check the action of $S$-duality on these classes of observables. $S$-duality is a generalization of the electro-magnetic duality and ${\cal N}$=4 SYM theory is conjectured to be invariant under it \cite{Montonen:1977sn,WittenOlive,Osborn:1979tq} . More specifically ${\cal N}$=4 SYM with gauge group $G$ and  generalized coupling constant $\tau$ is believed to be equivalent to the ${\cal N}$=4 SYM theory with the dual gauge group $^LG$ \cite{GNO}  and  coupling constant $^L\tau$

\begin{equation}^L\tau = -\frac{1}{n_g \tau}\end{equation}
\noindent
with $n_g=1, 2, 3$ depending on the choice of the gauge group. Since $S$-duality maps electric onto magnetic degrees of freedom, it establishes a natural isomorphism between operators. Explicit checks of the conjectured have been made for the action of the duality on chiral primary operators \cite{A,B,C}, surface operators \cite{D,E} and domain walls \cite{F,G}. For what concerns loop operators a nice calculation has been done in \cite{Okuda} where the prediction of the duality has been shown to hold to the next to leading order in the coupling constant expansion. Further and more general tests have been presented in \cite{Gomis:2011pf}, taking advantage of the exact results obtained from localization.\newline
Since different classes of electric observables preserving less supersymmetry are usually defined in ${\cal N}$ = 4 SYM theory (\emph{e.g.} Zarembo and DGRT Wilson Loops) it is interesting to understand the properties of their magnetic counterparts and how $S$-duality acts on them. In this paper we will try indeed to define and investigate new classes of 't Hooft operators preserving less supersymmetry than the circular and the straight line 't Hooft loops.\newline
Furthermore in ${\cal N}$ = 4 SYM theory there are also BPS mixed Wilson-'t Hooft loops operators which source both electric and magnetic \cite{TAKI,Kapustin:2005py,Kapustin:2006pk,Saulina:2011qr}\cite{Moraru:2012nu}: one can describe such operators requiring that the fields have a singularity near the loop as in (\ref{conf}) and inserting into the path-integral a factor

\begin{equation}\langle W(C) \rangle = \frac{1}{{\rm dim}(R)} {\rm Tr}_R \left[{\cal P}\exp\left(i \oint_c A_{\mu} dx^{\mu}\right)\right],\end{equation}
\noindent
 where $R$ is an irreducible representation of $G_B$ the stabilizer of $B$ \cite{Kapustin:2006pk}. These mixed operators are thus labeled by a pair ($B$, $R$) with $B$ a magnetic weight and $R$ a irrep. of $G_B$. It would also be interesting to investigate these dyonic operators since they are a very rich laboratory on which one can study the properties of $S$-duality.\newline
The plan of the paper is the following: in Section 2, as starting point, we define mixed BPS loop operators in an ${\cal N}$ = 4 Maxwell theory. We will analyze the supersymmetric properties of the singular configurations, we will calculate exactly their expectation value and show how the abelian $S$-duality acts on them. Section 3 is devoted to the analysis of 't Hooft loops in ${\cal N}$=4 SYM and in particular we define the magnetic counterpart of the Zarembo Wilson loop \cite{Zarembo:2002an} and the DGRT Wilson loop \cite{Drukker:2007qr}. In principle, using the formulae derived in that section, given a contour $C$ and the scalar couplings $\theta^A$ that define a supersymmetric Wilson loop in ${\cal N}$=4 SYM (see \cite{Dymarsky:2009si} for the detailed classification), we would be able to obtain the dual BPS 't Hooft operator. In Section 4 we calculate the expectation value for a particular example of magnetic operator up to one loop order, checking the action of $S$-duality on this non-trivial function of the coupling constant. The last section is dedicated to the conclusions and to present some open problems and directions for future investigations. There are four appendices : in the first one we summarize the notation, in the second one we discuss the cancellation at one-loop level of the fermion and boson excitation spectra of the non-zero modes around some 't Hooft backgrounds, in the third one some technical details about the integration over the adjoint orbit of $B$ are shown and in the last one some explicit configurations associated to 1/4 BPS 't Hooft operators are given.

\section{Wilson - 't Hooft loops in ${\cal N}$ = 4 supersymmetric Maxwell theory}      \label{se.2}

\subsection{Singular Configurations}
The starting point of our analysis is the ${\cal N}$ = 4 supersymmetric Maxwell theory. It contains a free photon, four free Weyl fermions and six neutral scalars. In this theory we will show how to  construct a
 wide  family of supersymmetric Wilson - 't Hooft loops. Depending on the specific form of the singular configurations these operators will preserve a number of super-symmetries which ranges from sixteen to
 two. In order to construct these objects, we have first to solve the classical Maxwell equations in the
 presence of a dyonic charge moving  along a closed non-selfintersecting curve $C$. In Feynman gauge,
 the  solution for the gauge potential can be easily expressed in terms of a contour integral
\begin{equation}
A_{\mu}(y) = \frac{\lambda}{2 \pi} \int_0^{2\pi} ds \frac{\dot{x}_{\mu}(s)}{(y-x(s))^2}\label{g}.\end{equation}
\noindent
In \eqref{g} the overall constant $\lambda$ is given by
\begin{equation}\lambda = g^2 \left(n + m \tau\right) = g^2 \left( n + m\frac{\theta}{2 \pi}\right) + 4 \pi m\, i \label{landa}\end{equation}
\noindent
where $n,m$ in $\mathbb{Z}$ are respectively the electric and the magnetic charges of the dyon and  $\tau$ is the complexified coupling constant (\ref{tau}); the functions $x_{\mu}(s)$ parameterize the closed circuit $C$.  If one evaluates   the {\it generalized} field strength
\begin{equation}F_{\mu \nu} = \underbrace{i\, \mathrm{Re}\left[ \partial_{\mu} A_{\nu} - \partial_{\mu} A_{\nu}\right]}_{\text{electric part}} + \underbrace{\mathrm{Im} \left[\frac{1}{2}\, \epsilon_{\mu \nu \rho \sigma}\, ( \partial_{\rho} A_{\sigma} - \partial_{\sigma} A_{\rho}) \right]}_{\text{magnetic part}}\label{fst}\end{equation}
\noindent
of the gauge connection \eqref{g},  one can immediately  check that it  correctly  describes  the electro-magnetic fields of a $(n,m)$ dyon and it possesses the  correct  singularity when approaching  the loop $C$. In
fact, in a  small neighborhood of a point $\overline{x} \in C$,  the circuit  in \eqref{g} can  be  approximated by the  straight line $x_{\mu}(s) \simeq \overline{x}_{\mu} + {\dot x}_{\mu}^0 s$ and  the behavior of \eqref{fst} for small $r$ reads
\begin{equation}F_{0 i} \sim  i n g^2 \frac{r^i}{r^3}\, \, \hspace{1cm} \, F_{i j} \sim  m \epsilon_{ijk}
\frac{r^k}{r^3}. \label{dmm}\end{equation}
\noindent
Here $r = | y - x |_{ _{\perp}}$ is the distance of the point $y$ from the straight line and the coordinate $i,j,k$ are transverse to the
straight line.  In order to define a BPS loop  operator  we have also to turn on the  scalar fields, by introducing
a sort of $R-$symmetry current proportion to six-component vector $\theta^{A}(s)$. The solution of the equations of motion  for the fields $\phi^{A}$ can be easily determined and they take the form
\begin{equation}\phi(y)^A = \frac{|\lambda|}{2 \pi} \int ds \frac{\theta^A(s)}{(y-x(s))^2}, \label{f4}\end{equation}
\noindent
where $A=1\dots 6$ and the scalar couplings $\theta^A$ are taken to obey  the  standard constraint $\theta^A \theta^A = 1$.   The next step  is to determine under which conditions the field configurations  \eqref{g} and \eqref{f4} define a loop operator which preserves  a certain amount of supersymmetries.

\subsection{Supersymmetric properties}
The $U(1)$ gauge connection \eqref{g} and  the scalar fields \eqref{f4} define a BPS operators
if they annihilate the supersymmetry transformation of the fermion field $\Psi$, {\it i.e.}

\begin{equation}\delta \Psi =\frac{1}{2}\, \Gamma^{MN} F_{MN}\, \epsilon - 2 \Gamma^A \phi^A \epsilon_1 = 0 \label{dmmm}\end{equation}
\noindent
where $M,N= 0,\dots,9, $ and $A=4,\dots,9$ and we used  the  usual ten dimensional notation (see app. A for additional details on our conventions). The parameter $\epsilon = \epsilon_0 + x_{\mu} \Gamma^{\mu} \epsilon_1$ in \eqref{dmmm} is a conformal Killing spinor
in $\mathbb{R}^4$ since  $\epsilon_0$ and $\epsilon_1$ are  two
constant  Majorana-Weyl  spinors of opposite  chirality.  The former generates  the Poincar\`{e} sector while  the latter is responsible for the  conformal supersimmetries.  The condition \eqref{dmmm} can be translated into an algebraic local  constraint, which contains only the circuit and the coupling $\theta^{A}$.\newline
In  fact let us consider the supersymmetric variation of the gaugino $\Psi$ for the classical configurations \eqref{g} and   \eqref{f4}:

\small
\be
\label{pp}
\delta \Psi = |\lambda| \oint ds \left( \frac{(y-x)_{\mu}}{(y-x)^4} \left[- 2 i \Gamma^{\mu \nu}\,  \cos{\varphi}\, \dot{x}_{\nu} - \epsilon_{\mu \nu \rho \sigma} \Gamma^{\rho \sigma}\,  \sin{\varphi}\, \dot{x}_{\nu} - 2 \Gamma^{\mu A} \theta^A\right] \epsilon(y)- 2 \frac{\Gamma^A \theta^A}{(y-x)^2}\, \epsilon_1 \right),
\ee
\normalsize
\noindent
where $|\lambda|$ and  $\varphi$  are the modulus and  the phase of the complex number $\lambda$ and
$ \epsilon(s)\equiv\epsilon_0 + \Gamma^{\nu}x_{\nu}(s)\epsilon_1$.  By adding and subtracting the same
term to eq. \eqref{pp}, it can be rearranged as follows
\begin{equation}
\begin{split}
\delta \Psi =&  |\lambda| \oint ds \left( \frac{(y-x)_{\mu} \Gamma^{\mu}}{(y-x)^4} \left[ i \Gamma^{\nu}\,  \cos{\varphi}\, \dot{x}_{\nu} + \Gamma^{1234} \Gamma^{\nu} \sin{\varphi}\, \dot{x}_{\nu} + \Gamma^{A} \theta^A\right] ( \epsilon_0 + \Gamma^{\nu}x_{\nu} \epsilon_1 )  \right.-\\
& \left. -2 \Gamma_{\mu} \frac{(y-x)_{\mu}}{(y-x)^4} \left[ i \Gamma^{\nu}  \dot{x}_{\nu} + \Gamma^{A}\theta^A\right] \Gamma^{\rho}(y-x)_{\rho} \epsilon_1 - 2  \frac{ \Gamma^A \theta^A}{(y-x)^2}\, \epsilon_1    \right)
\end{split}
\label{var1}
\end{equation}
The second line in eq. (\ref{var1}) can be shown to  vanish identically by exploiting a little bit of {\it Diracology} and a trivial integration by parts. In the first line we have used the identity
\begin{equation}\epsilon_{\mu \nu \rho \sigma} \Gamma^{\rho \sigma} = - 2 \Gamma^{\mu \nu} \Gamma^{1234},\end{equation}
in order to eliminate the Levi-Civita tensor from \eqref{var1}. Therefore our configuration is supersymmetric
only when
\begin{equation}\delta \Psi = \oint ds \left( \frac{(y-x)_{\mu} \Gamma^{\mu}}{(y-x)^4} \left[ i \Gamma^{\nu}\,  \cos{\varphi}\, \dot{x}_{\nu} + \Gamma^{1234} \Gamma^{\nu} \sin{\varphi}\, \dot{x}_{\nu} + \Gamma^{A} \theta^A\right] ( \epsilon_0 + \Gamma^{\nu}x_{\nu} \epsilon_1 )  \right) = 0.
\label{pp3}
\end{equation}
Eq. \eqref{pp3} is, in turn, zero \emph{iff}\footnote{In order to see that we can note that if eq. ($\ref{pp3}$) is equal to zero also the following expression is vanishing

\begin{displaymath}\Gamma^{\alpha} \partial_{\alpha} \oint ds \left( \frac{(y-x)_{\mu} \Gamma^{\mu}}{(y-x)^4} \left[ i \Gamma^{\nu}\,  \dot{x}_{\nu} + \Gamma^{A} \theta^A\right] ( \epsilon_0 + \Gamma^{\nu}x_{\nu} \epsilon_1 )  \right) = 0
\end{displaymath}
\noindent
that is equal to

\begin{displaymath}\oint ds\, \,  \delta (y-x)\left[ i \Gamma^{\nu}\,  \dot{x}_{\nu} + \Gamma^{A} \theta^A\right] ( \epsilon_0 + \Gamma^{\nu}x_{\nu} \epsilon_1 ) = 0
\end{displaymath}
\noindent
and since this equation must hold for every value of the variable $y$ one can choose $y=x(s)$ and integrating over $s$ one obtains exactly the equation ($\ref{dionici}$).} the integrand vanishes
\begin{equation}
\left[ i \Gamma^{\mu}\,  \cos{\varphi}\, \dot{x}_{\mu} + \Gamma^{1234}\Gamma^{\mu}\, \sin{\varphi}\, \dot{x}_{\mu} +  \Gamma^{A} \theta^A\right] ( \epsilon_0 + \Gamma^{\nu}x_{\nu} \epsilon_1 ) = 0.\label{dionici}
\end{equation}
This is the advertised local constraint which determines the BPS nature of our loop operator only in terms of
the   the scalar couplings $\theta^A(s)$ and the circuit $x_{\mu}(s)$. For a merely electric loop, {\it i.e.}
$\varphi=0$,  eq. \eqref{dionici} takes the simplified form
\be
\label{pippo}
\left[ i \Gamma^{\mu}\,   \dot{x}_{\mu} +  \Gamma^{A} \theta^A\right] ( \epsilon_0 + \Gamma^{\nu}x_{\nu} \epsilon_1 ) = 0,
\ee
which is the usual BPS condition for a Maldacena-Wilson operator in $\mathcal{N}=4$. For $\varphi\ne0$
we can define
\begin{equation}
\epsilon^{\prime}(x) = e^{-i \Gamma^{1234} \frac{\varphi}{2}} \epsilon(x)\label{epsilonmapping},
\end{equation}
which is obtained through a four-dimensional chiral rotation of the original spinor. This auxiliary quantity
again obeys \eqref{pippo}, {\it i.e.} the classification of the dyonic abelian loop operators reduces to that of the ordinary BPS Wilson loops.\newline
The general solution of \eqref{pippo} was obtained in \cite{Dymarsky:2009si}.
  There  the key step was to recast \eqref{pippo} in a covariant ten dimensional language  by introducing the vector  $v^M =\left \{\frac{dx^{\mu}}{ds}, \theta^I(s)\right\}$.
One finds
\begin{equation}
v^M(x) \Gamma_M \epsilon^{\prime} (x)= 0 \label{WL2}
\end{equation}
where $\Gamma_{M} = (\Gamma_{\mu}, \Gamma_I)$ denotes, as usual,   the ten  dimensional Dirac matrices.  Then to solve the above linear system,  one considers $\epsilon^{\prime}$  as given and looks for the couplings $v^M$ which obey \eqref{WL2}. One can distinguish two different families of solutions depending on whether or not the vector $u^M = \epsilon \Gamma^M \epsilon$ identically vanishes.

\noindent
When $u^M \neq 0$, there is a unique solution of eq. (\ref{WL2}) and it is given by $v^M=\kappa u^M$ with $\kappa$ a complex number \cite{Dymarsky:2009si}. The resulting loops are the orbits of the conformal transformations generated by $Q^2_{\epsilon^{\prime}(x)}$. If we consider only closed loops, we obtain operators defined on $(p,q)$ \emph{Lissajous figures} where $p$ and $q$ are integer numbers. The quantum
properties for these operator in the pure  electric case were studied in \cite{Cardinali:2012sy}.

\noindent
If  $u^{M}$ vanishes  identically
  on a submanifold $\Sigma_{\epsilon} \subseteq \mathds{R}^{4}$, $\epsilon$
  is a \textit{pure spinor} on $\Sigma_{\epsilon}$ and consequently it induces an almost complex structure $J_{\epsilon}$ on this region
  \cite{Dymarsky:2009si}.    The possible  solutions  $v^{M}$ of eq. \eqref{WL2} in a point $x\in \Sigma_{\epsilon}$  are   then provided by all the anti-holomorphic  vectors with respect to $J_{\epsilon}$   \cite{Dymarsky:2009si}.  This result can be used to associate a supersymmetric loop operator to each  closed contour $\gamma$ in $\Sigma_{\epsilon}$. An explicit  construction   of the vector $v^{M}$, modulo equivalence under the action of the superconformal group, for the possible choice of
  $\Sigma$  can be found in \cite{Dymarsky:2009si}.

  \noindent
Thus we have shown that if we define a dyonic operator as in (\ref{g},\ref{f4}) with the configurations supported on a path $C$ and characterized by the scalar coupling $\theta^A(s)$ in such a way that ($C$, $\theta^A(s)$) describe a supersymmetric Wilson loop, the mixed eletric-magnetic configurations obtained preserve the same amount of supersymmetry of the electric operators. Moreover the supercharges preserved by the two classes of observables are not the same but are related by a peculiar trasformation that can be read from (\ref{epsilonmapping}).

\noindent
What we have learnt is not surprising, rather it is what $S$-duality predicts.  Indeed it has already been shown in the literature \cite{WittenOlive,Olive:1995sw} that while all bosonic symmetry generators are mapped trivially under the Montonen-Olive duality, the supersymmetry generators are multiplied by a $\varphi$-dependent phase:

\begin{equation}\bar{Q}_{\dot{\alpha}} \rightarrow e^{i \frac{\varphi}{2}}\bar{Q}_{\dot{\alpha}} \, \, \, \, Q_{\alpha} \rightarrow e^{- i \frac{\varphi}{2}} Q_{\alpha}\end{equation}
\noindent
depending on the chirality of the spinor. The physical reason is that the supersymmetry algebra in presence of a central extension is modified as

\begin{eqnarray}
\nonumber  \{ Q_L, Q_L \} \sim i q + g = |\lambda| \left( i \cos{\varphi}  + \sin{\varphi} \right)\\
  \{ Q_R, Q_R \} \sim i q - g = |\lambda| \left( i \cos{\varphi}  - \sin{\varphi} \right)
\label{ho1}\end{eqnarray}
\noindent
with $q$ and $g$ the electric and the magnetic charges of the configuration. It's thus natural that after an electric-magnetic transformation the supercharges acquire a chirality dependent phase equal to $e^{\pm i\varphi/2}$.

\subsection{Expectation values of the operators}
Above  we have constructed new BPS dyonic configurations. The next step is the calculation of their expectation value and we can do that by firstly evaluating
the classical action
\begin{equation}
\label{pippa}
S_{{\cal N}=4 Max}^0 = \frac{1}{2 g^2}  \left( \int_{R^4} \frac{1}{2}F_{\mu \nu} F^{\mu \nu}  +  D_{\mu} \phi^A  D^{\mu} \phi^A  \right) - i \frac{\theta}{32 \pi^2} \left( \int_{R^4} F_{\mu \nu} \tilde{F}^{\mu \nu}\right)\end{equation}
\noindent
on the field configurations (\ref{g},\ref{f4}). It is easy to calculate separately the three different contributions in \eqref{pippa}. The scalar part reads
\begin{equation}\frac{1}{2 g^2}\int d^4y\, D_{\mu} \phi^A D^{\mu} \phi^A =  \frac{\lambda^2}{2 g^2}\, S_{12} , \end{equation}
while the  gauge contribution and  the theta term are respectively  given by
\begin{equation}
\begin{split}
\frac{1}{4\, g^2}\int d^4y\,  F_{\mu \nu}  F^{\mu \nu} =& - \frac{1}{2} \left[ g^2 \left(n+\frac{m \theta}{2 \pi}\right)^2 - \frac{4 \pi^2 m^2}{g^2} \right] G_{12}
\hspace{1cm}\\
-i \frac{\theta}{32 \pi^2} \left( \int_{R^4} d^4y\, F_{\mu \nu} \tilde{F}^{\mu \nu} \right)= &\, \frac{g^2}{2} \left[  \frac{n m}{\pi} + \frac{m^2 \theta^{\, 2}}{4 \pi} \right] G_{12}
\end{split}
\end{equation}
The function $S_{12}$ and $G_{12}$  are defined as the following  {\it formal} contour integral\footnote{Separately the function $S_{12}$ and $G_{12}$ are ultraviolet
divergent.}
\begin{equation}S_{12} = \frac{1}{4 \pi^2} \oint ds\, dt\, \frac{\theta^A(s) \theta^A(t)}{(x(s)-x(t))^2} \hspace{1cm} G_{12} = \frac{1}{4 \pi^2} \oint ds\, dt\, \frac{\dot{x}(s) \dot{x}(t)}{(x(s)-x(t))^2} .\label{G12S12} \end{equation}
\noindent
Collecting all the contributions we obtain that the on-shell action evaluated on the classical configuration can be written as
\begin{equation}S_{{\cal N}=4 Max}^0 = -\frac{g^2}{2} \left[ n^2 - \left( \frac{m \theta}{2 \pi}\right)^2 - \frac{4 \pi^2 m^2}{g^4} \right] G_{12} + \frac{g^2}{2} \left[ \left( n + \frac{m \theta}{2 \pi}\right)^2 + \frac{4 \pi^2 m^2}{g^4} \right] S_{12}  \end{equation}
\noindent
However this is not the end of the story since, in order  to calculate the $VEV$ of the dyonic operator, we have also to insert in the path integral a Wilson loop of the form \cite{Kapustin:2005py}
\begin{equation}W(C) = \exp\left[  - \frac{1}{2 \pi}\oint \left(  n\, \text{Re}[A_{\mu} \dot{x}^{\mu}] - \frac{|\lambda|}{g^2} \phi^A \theta^A  \right) ds\right]\end{equation}
\noindent
and to evaluate also  its value  on the classical configurations. 
Since the insertion of the operator contributes as

\begin{equation} g^2 \left[- n^2 -  \frac{n m \theta}{2 \pi} \right] G_{12} +    g^2 \left[ \left( n + \frac{m \theta}{2 \pi}\right)^2 + \frac{4 \pi^2 m^2}{g^4} \right] S_{12}\end{equation}
\noindent
the final expression is given by

\begin{equation}\langle W\, H\rangle =  \text{exp}\left[- \frac{\lambda^2}{2\, g^2} \left( G_{12} - S_{12} \right)\right].\end{equation}
\noindent
This result is exact because we are dealing with a Gaussian theory and finite for locally BPS operators on smooth contour since a nice cancellation occurs when
the variables  $s$ and $t$ coincide during the integration  \cite{Drukker:1999zq}.

\section{'t Hooft Loops in $U(N)$ ${\cal N}$=4 SYM}

\subsection{Introduction}
This section is devoted to extend the abelian construction of the loop operators presented in the previous chapter to the non-abelian
${\cal N}$=4 SYM theory. More precisely we will consider the case in which the gauge group is $G$ =  $ ^L G$ = $U(N)$ leaving the extension to a generic group for a future investigation. In such theory we will define two large families of magnetic operators that are dual to Zarembo and DGRT Wilson loops \cite{Zarembo:2002an}\cite{Drukker:2007dw}.\newline
In order to construct them, following the standard procedure presented in \cite{Kapustin:2005py}, we have to embed the abelian construction into the non-abelian group $G$ defining an homomorphism

\begin{equation}\rho\, :\,  U(1) \rightarrow G.\end{equation}
\noindent
The most general homomorphism $\rho$ maps $e^{i \alpha} \in U(1)$ to the the diagonal matrix
\begin{equation}G = \text{exp}( i\alpha B) = \text{diag}(\, \text{exp}\,(i m_1 \alpha),\, \text{exp}\,(i m_2 \alpha), . . . , \text{exp}\,(i m_N \alpha))  \end{equation}
\noindent
 with the $N$-plet of integers $^L w = (m_1,m_2, . . . ,m_N)$ that identifies a coweight vector (magnetic weight) of $G$.
 The magnetic weights of $U(N)$ are in one-to-one correspondence with the Young tableaus containing $m_l$ boxes in the l-th row and identify an irreducible representation of the group. As example if we choose to define the 't Hooft observable in the fundamental, the $k$-symmetric or the $k$-antisymmetric representation of $U(N)$ the diagonal matrix $B$ will be respectively of the form

 \small
\begin{equation}B_{\text{F}} = \text{diag} \left( \underbrace{1, 0, 0 ......0}_N \right),\, B_{\text{ k-sym}}= \text{diag} \left( \underbrace{k, 0, 0 ......0}_N \right),\, B_{\text{ k-ant}} = \text{diag} \left( \underbrace{1, 1, ......1}_k, \underbrace{0,....0}_{N-k} \right).\end{equation}

\normalsize
\noindent
The singular gauge configurations associated to the magnetic operator can thus be constructed as

\begin{equation} A_{\mu\,}(y) =  B\, \int ds \frac{\dot{x}_{\mu}(s)}{(y-x(s))^2}\label{gi1}\end{equation}
\noindent
and using the same embedding one can define the scalar configurations

 \begin{equation} \phi^{A}(y) =  B\, \int ds \frac{\theta^A(s)}{(y-x(s))^2}. \label{f}\end{equation}
\noindent
At this point it's not difficult to perform the supersymmetric variation of the non-abelian configurations

\begin{equation}\delta_{\epsilon} \Psi = \frac{1}{2}\Gamma^{MN} F_{MN}\, \epsilon(s) - 2 \phi^A \Gamma^A \epsilon_1 = 0\end{equation}
\noindent
and note that, since the color structure factorizes out, we obtain the same condition on the path $C$ and the scalar couplings $\theta^A(s)$ found in the abelian case. As consequence the magnetic configuration defined as in (\ref{gi1}, \ref{f}) with the same ($C$, $\theta^A(s)$) of a BPS Wilson loop will preserve the same number of supersymmetries of the electric observable.\newline
In the following subsections we will show explicitly the singular configurations associated to two classes of BPS 't Hooft operators and verify their supersymmetric properties. More in detail we will analyze the straight line and the circular operator and provide their respective generalizations, \emph{i.e.} Zarembo and DGRT 't Hooft loops.

\subsection{1/2 BPS 't Hooft line}

Let us begin considering the simplest case of 't Hooft loop operator namely when the observable is defined on a straight line. Without loss of generality we can parameterize the circuit and the scalar couplings as follow

\begin{equation}x_{\mu} =(\, 0, 0, 0, s ), \, \, \, \, \, \theta^A(s) = \theta_0^A \label{straightline}\end{equation}
\noindent
with the variable $s$ ($-\infty < s < +\infty$) that describes the circuit,  $\mu=1..4$, $A=0,5..9$ and  $\theta_0^A$ a six-dimensional constant vector that satisfies $\theta_0^A \theta_0^A$ = 1. It's easy to see from (\ref{gi1}, \ref{f}) that the associated configurations are given by

\begin{equation}A_{4}(y) =   \frac{B\, \pi}{|\, y_{\bot} |} \hspace{1cm} \phi(y)^A =  \frac{\pi\, B\, \theta_0^A}{|\, y_{\bot} |}\label{line1}\end{equation}
\noindent
where $y_{\perp}^2 = y_1^2 + y_2^2 + y_3^2$ and the other components of $A_{\mu}$ ($\mu = 1..3$) are identically zero. In order to show that  these configurations preserve a part of the supersymmetries we have to verify that the following variation

\begin{equation}\delta \Psi = \frac{1}{2} \Gamma^{\mu \nu} F_{\mu \nu} (\epsilon_0 + x_{\mu} \Gamma^{\mu} \epsilon_1 ) + \Gamma^{\mu A} \partial_{\mu} \phi^A (\epsilon_0 + x_{\mu} \Gamma^{\mu} \epsilon_1 ) - 2 \tilde{\Gamma}_A \phi^A \epsilon_1 \label{equation}\end{equation}
\noindent
vanishes identically when it is evaluated on (\ref{line1}). After some manipulations one can shows that it reduces to one independent constraint for the spinors $\epsilon_0$ and $\epsilon_1$

\begin{equation}\left( \Gamma^{123} + \Gamma^A \theta_0^A\right)\epsilon_0 =0\, \, \, \, \hspace{1cm} \left(  \Gamma^{123}  - \Gamma^A \theta_0^A\right)\epsilon_1 =0.\, \, \, \,\label{magneticooo}\end{equation}
\noindent
Since the matrices $\left(  \Gamma^{123}  \pm \Gamma^A \theta_0^A\right)$ are nilpotent we can easily conclude that the operator preserves half of the super Poincar\'{e} plus half of the super-conformal charges of the theory.\newline
Now we want to investigate the relation between these supercharges and those preserved by the Wilson operator defined as

\begin{equation}\langle W(C)\rangle = \frac{1}{{\rm dim}(R)} {\rm Tr}_R\left[ {\cal P} \exp\left( \oint_C \left( i A_{\mu} \dot{x}^{\mu} + \phi^A \theta^A(s) |\dot{x}|\right) ds \right)\right],\label{MaldacenaWilsone}\end{equation}
\noindent
where the circuit $C$ and the scalar couplings are the same as in (\ref{straightline}). To do that we have to compare the equations (\ref{magneticooo}) with the supersymmetric variation of (\ref{MaldacenaWilsone}) given by

\begin{equation}\left(  i \Gamma^{4}   + \Gamma^A \theta_0^A\right)\epsilon_0 =0\, \, \, \,\hspace{1cm} \left(  i \Gamma^{4}   - \Gamma^A \theta_0^A\right)\epsilon_1 =0.\, \, \, \,\label{electric}\end{equation}
\noindent
It is not difficult to check that, as expected from $S$-duality consideration (see previous section), the spinors $\epsilon_0^m$ and $\epsilon_1^m$ satisfying eq.(\ref{magneticooo}) are related to the solutions of eq.(\ref{electric}) $\epsilon_0^e$ and $\epsilon_1^e$ by the four dimensional chiral rotation

\begin{equation}\epsilon_{\, 0}^{\, \, m} = \text{exp}\left( \frac{i}{2} [\Gamma^{1234} ]\, \frac{\pi}{2} \right) \epsilon_0^e \, \, \, \,\hspace{1cm} \epsilon_{\, 1}^{\, \, m} = \text{exp}\left( -\frac{i}{2} [\Gamma^{1234} ]\, \frac{\pi}{2} \right) \epsilon_1^e . \label{BO}\end{equation}

\subsection{Zarembo 't Hooft loop}
A simple idea to solve the BPS equation for the Wilson operator given by

\begin{equation}\left( i \Gamma_{\mu} \dot{x}_{\mu}(s) + \Gamma^A \theta^A(s) \right)( \epsilon_0 + x_{\nu}(s) \Gamma^{\nu} \epsilon_1 ) = 0\label{23}\end{equation}
was proposed in \cite{Zarembo:2002an} by Zarembo. For an arbitrary shape of the loop $C$ the author chose the scalar couplings $\theta^A(s)$ in such way that the resulting operators preserve at least two supercharges. If

\begin{equation}\theta^A(s)\, = M_{\mu}^A \dot{x}_{\mu}(s)\label{newcoupling}\end{equation}
\noindent
with $M_{\mu}^A$ a rectangular 4 $\times$ 6 matrix that satisfies $M_{\mu}^A M_{\nu}^A $ = $\delta_{\mu \nu}$, he showed that, considering only the super-Poincar\'{e} trasformation,  all the dependence from the circuit is totally dropped out in the equation (\ref{23}). Only the dimensionality of the subspace in which the curve lies is important to determine the number of supercharges. Indeed the equation (\ref{23}) can be rewritten as

\begin{equation} \left( \Gamma_{\mu} - i \Gamma^A M_{\mu}^A \right) \epsilon_0 = 0\label{24}\end{equation}
\noindent
and one can see that, since $\mu$ ranges from one to four and since each equation halves the number of preserved supercharged, for a generic contour the operator is 1/16 BPS.\newline
Following the Zarembo's idea we can extend the magnetic line operator (\ref{line1}) and construct the Zarembo 't Hooft loops. They are defined by specify the singular configurations associated to the operators as

\begin{equation}A_{\mu}(y) = B \int ds \frac{\dot{x}_{\mu}}{(y-x(s))^2}\, \, \hspace{1cm} \phi(y)^A =  B \int ds \frac{M_{\mu}^A \dot{x}_{\mu}}{(y-x(s))^2}. \end{equation}
\noindent
\normalsize
If we consider only the super-Poincar\'{e} transformations and choose a generic curve inside $\mathbb{R}_4$, the supersymmetric variation of the magnetic observables is given by four independent equations
\normalsize
\begin{equation} \left(\epsilon_{\mu \nu \rho \sigma} \Gamma^{\nu \rho \sigma} - \Gamma^A M_{\mu}^A\right)\epsilon_0 =0\label{dfwf}\end{equation}
\noindent
and as consequence these operators are at least 1/16 BPS. However, as in the electric case if the curves lies on a subspace of $\mathbb{R}_4$ an enhancement of the supersymmetry occurs. If we compare the variation of the Zarembo Wilson operator to the eq. (\ref{dfwf}) we can note again that the preserved supercharges are not the same but, as previously underlined, are related by the transformation (\ref{BO}). Finally in order to know how many super-conformal charges are preserved one has to give the explicit shape of the loop, however for a generic circuit no super-conformal transformation will be preserved neither for the electric nor for the magnetic observable.\newline
In the appendix (D.1) we will present an explicit configuration for a 1/4 BPS magnetic loop of Zarembo-type, \emph{i.e.} an operator supported on a cusp at angle $\alpha$, and we will study its supersymmetric properties.

\subsection{1/2 BPS Circular 't Hooft Loop}\label{maxim}

The circular 't Hooft loop operator is already well known in the literature since the work of Kapustin \cite{Kapustin:2005py} and here we only review briefly its construction using our notation. Let us start choosing the couplings that identify the operator as follow

\begin{equation}x_{\mu}(s)=( \cos{(s)}, \sin{(s)}, 0, 0), \, \, \, \hspace{1cm} \, \, \theta^A(s)\, = \theta^A_0\, \, \hspace{0.5cm} \end{equation}
\noindent
where $s$ ranges from zero to $2\pi$. The classical configurations associated to the observable and obtained from (\ref{gi1}) and (\ref{f}) read as
\small
\begin{equation}A_{1}(y) =  2 \pi B \frac{y_2 \left( 1 + y^2 - \sqrt{(1+y^2)^2 - 4 y_{\|}^2}\right)}{y_{\|}^2\, \, \left( \sqrt{(1+y^2)^2 - 4 y_{\|}^2}\right)}\, \, \, \, \, \, A_{2}(y) = 2 \pi B \frac{y_1 \left( 1 + y^2 - \sqrt{(1+y^2)^2 - 4 y_{\|}^2}\right)}{y_{\|}^2\, \, \left( \sqrt{(1+y^2)^2 - 4 y_{\|}^2}\right)}\end{equation}

\begin{equation}\phi^A(y) = 2 \pi B \frac{\theta^A_0}{ \left( \sqrt{(1+y^2)^2 - 4 y_{\|}^2}\right)} \label{F}\end{equation}
\normalsize
\noindent
where $y_{\|}^2=y_1^2 + y_2^2$ and $y^2= y_1^2+y_2^2+y_3^2+y_4^2$. The configurations found can be obtained from those presented in \cite{PestunII,Gomis:2009xg,Kapustin:2005py} and usually defined in $AdS_2 \times S^2$ by a simple conformal transformation that maps $\mathbb{R}^4$ to $AdS_2 \times S_2$  (see appendix A of \cite{Okuda} for the explicit form of the mapping\footnote{In $AdS_2 \times S^2$ the subgroup of the conformal group preserved by the operator acts as isometries and the explicit form of the fields is simpler.}). The supersymmetric variation of the operator can be written as

\begin{equation}(    \Gamma^{34} \epsilon_1 + \Gamma^{A} \theta^A_0 \epsilon_0)=0 \hspace{1cm}(\text{Circular 't Hooft loop}) \end{equation}
\noindent
and compared to the variation of the ordinary electric operator
\begin{equation}( i \Gamma^{12} \epsilon_1 - \Gamma^{A} \theta^A_0 \epsilon_0)=0 \hspace{1cm}(\text{Circular Wilson loop}) . \end{equation}
\noindent
Both the operators preserve exactly one half of the supersymmetries of the theory and more precisely some linear combinations of the super-Poincar\'{e} and super-conformal charges.

\subsection{DGRT 't Hooft Loops}
\label{sectionDGRT}
Following the construction of the Wilson loops on $S^3$ introduced by Drukker \emph{et al.} \cite{Drukker:2007qr} we can define new BPS 't Hooft operators generalizing the 1/2 BPS circular magnetic observable presented in the previous subsection. Just for simplicity we present only the case in which the loop lies on a two sphere but no restriction arises if we want to extend the definition to a general loop on $S^3$.\newline
In order to construct these magnetic operators one has to choose the circuit as

\begin{equation}x_{\mu}(s) = (x_1(s),  x_2(s), x_3(s), 0)\, \, \, \, \, \, \, 0 < s < 2 \pi\end{equation}
\noindent
with $x_1^2+x_2^2+x_3^2=1$ and the scalar couplings as
\begin{equation}\theta^A(s) = M_K^A \epsilon^{IJK} \dot{x}_I(s)\, x_J(s)\label{DGRT}\end{equation}
\noindent
where $M_K^A$ is a generic three by six matrix that satisfies $M_I^A M_J^A = \delta^{IJ}$. With this ansatz the singular configurations are given by

\begin{equation}A_I(y) =  B \int_0^{2\pi} ds \frac{ \dot{x}_I}{(y-x(s))^2}\, \, \, \, \hspace{0.5cm} \phi(y)^A = B \int_0^{2\pi} ds \frac{M_K^A \, \epsilon^{IJK} \dot{x}_I(s) x_J(s)}{(y-x(s))^2}. \label{II}\end{equation}
\noindent
To obtain the explicit configuration and study its supersymmetric properties one has to know the path $C$ on which the operator is supported.  However it's not difficult to show that for a generic curve inside the two sphere the singular configurations (\ref{II}) define an 1/8 BPS object.\newline
In the appendix (D.2) and (D.3) two types of BPS DGRT magnetic operators will be presented in detail : the first is supported on a latitude at polar angle $\theta_0$ and the second on a wedge, a loop made of two arcs of length $\pi$ connected at an arbitrary angle $\delta$, \emph{i.e.} two longitudes on the two-sphere. Their supersymmetric properties will be carefully analyzed and compared to those of the dual electric observables .

\section{'t Hooft loops expectation value}
\subsection{Introduction}

In this section we compute explicitly the expectation value of some BPS 't Hooft operators up to
next-to-leading order in perturbation theory generalizing the calculation presented in \cite{Okuda} for the 1/2 BPS circular loop. In a future investigation it could be really intriguing to go beyond the perturbative analysis deriving an exact expression for the $VEV$ of these
magnetic observables using localization techniques.\newline
In the semiclassical approximation the $VEV$ of the 't Hooft operators is simply given by

\begin{equation}\langle\, H\, \rangle = \text{exp}^{-S_0},\end{equation}
\noindent
namely by the contribution to the ${\cal N}$ = 4 SYM action of the singular configurations associated to the operators

\begin{equation} A_{\mu\,}^0 =  B \int ds \int \frac{d^4p}{(2\pi)^4} \frac{\dot{x}_{\mu}(s)\, e^{i p(y-x(s))}}{p^2}\, \hspace{0.5cm} \, \, \phi_{0}^A =  B \int ds \int \frac{d^4p}{(2\pi)^4} \frac{\theta^A(s)\, e^{i p (y-x(s))}}{p^2},\label{jkl1}\end{equation}
\noindent
where as usual $B$ is a magnetic weight of $U(N)$, $x_{\mu}(s)$ is the circuit on which the operators are supported and $\theta^A(s)$ are the scalar couplings. Writing down the bosonic part of the ${\cal N}$ = 4 super Yang-Mills action as
\begin{equation}S_{{\cal N}= 4 } =  \frac{1}{2 g^2}  \int_{R^4} d^4x\, \text{tr} \left( F_{\mu \nu} F^{\mu \nu}  + 2 D^{\mu} \phi^A D_{\mu} \phi^A +  [ \phi^A, \phi^B]^2  \right)\label{pk},\end{equation}
\noindent
it's straightforward to evaluate it on the background (\ref{jkl1}) and find that

\begin{equation}S_0 = \frac{4\, \text{Tr}B^2\, \pi^2}{g^2} \left( G_{12} + S_{12} \right)  \end{equation}
\noindent
where $G_{12}$ and $S_{12}$ are defined in (\ref{G12S12})\footnote{The coefficient $\text{Tr}B^2$ is equal to 1 if the operator is in the fundamental representation, $k$ if it is in the $k$-antysymmetric irrep, $k^2$ if it is in the $k$-symmetric irrep and finally $\sum_{i}^N m_i^2$ for a generic representation.}. Since this value is divergent, in order to regularize it one has to introduced a counter-term defined on the hypersurface $\Sigma$, the boundary of a solid tubular neighborhood of the contour $C$ \cite{Okuda,PestunII}. Evaluating this term on the classical configurations one obtains

\begin{equation}S_{boundary} = - \frac{2 \pi^2 \text{Tr}B^2}{g^2} S_{12}\end{equation}
\noindent
that summed to the on-shell action makes correctly the regularized action convergent\footnote{As already noted in section two the combination $G_{12}-S_{12}$ is finite even if $G_{12}$ and $S_{12}$ are separately divergent.}

\begin{equation}\langle\, H\, \rangle = \text{exp}^{-S_0 - S_{boundary}} = \text{exp}\left[-\frac{4\, \text{Tr}B^2 \pi^2}{g^2} \left( G_{12} - S_{12} \right)\right].\end{equation}
\noindent
In order to go beyond and carry out the quantum computation we have to perform the path integral expanding the quantum fields around the singular configurations

\begin{equation}A^{\mu} = A_0^{\mu} + \hat{A}^{\mu}\, \, \, \hspace{0.5cm} \, \phi^A = \phi^A_0 + \hat{\phi}^A\end{equation}
\noindent
where ($\hat{A},\hat{\phi}$) are the non-singular quantum fluctuation
on which we have to integrate over \cite{Okuda}. To quantize the theory in the background of $A_0^{\mu},\phi^A_0$ we have also to fix the gauge imposing for example that

\begin{equation}D_{0}^{\mu} \hat{A}_{\mu} - i [ \phi^A_0, \hat{\phi}^A ] = 0\end{equation}
\noindent
and then add to the ${\cal N}$=4 SYM action the gauge fixing plus the ghost contributions that in a ten dimensional notation can be written as

\begin{equation}S_{gf+g} = \frac{1}{g^2} \int d^4x\, \text{tr}\left[ D_0^M \hat{A}_M D_0^N \hat{A}_N - \bar{c} D_0^M D_{M } c\right].\end{equation}
\noindent
At this stage the operator is gauge-dependent and thus we have to introduce a procedure to restore the gauge invariance. Following the reference \cite{Okuda} the idea it to include in the path integral definition the integration over the adjoint $G$-orbit of the magnetic weight $B$

\begin{equation}O(B) = \left[ B^g = g B  g^{-1}, g \in U(N)\right]\end{equation}
\noindent
which is diffeomorphic to the coset space $U(N)/H$ with $H$ the invariance group of the weight $B$. On $O(B)$ one defines a metric
and fixing carefully its normalization ( for more detail and convention see Appendix B) one can perform the integration over the adjoint orbit obtaining

\begin{equation}\int ds_{O(B)}^2 = \left(\frac{4 \pi^2}{g^2} ( G_{12} - S_{12} )\right)^{\text{dim}(U(N)/H)/2} \text{Vol}(U(N)/H) \prod_{ \alpha>0}\text{Tr}[E_{\alpha}, B]^2 \label{adjoint}\end{equation}
\noindent
where $E_{\alpha}$ are the ladder operators associated to roots $\alpha$ of the Lie
algebra $su(N)$ and the product is over the positive roots that don't belong to the invariance subgroup of $G$. Up to one loop order the expectation value of the 't Hooft loop is thus given by

\begin{eqnarray}
\nonumber  \langle\, H\, \rangle =
 \text{exp}\left[-\frac{4\, \text{Tr}B^2 \pi^2}{g^2} \left( G_{12} - S_{12} \right)\right] \frac{[ \text{det}( i \Gamma^M D_{0\,M})]^{1/4}\, \,\text{det}_g(-D_0^2)}{[\text{det}_b(-\delta^{MN}D_0^2 + 2 i F_0^{MN})]^{1/2}} \times \\
   \left(\frac{4 \pi}{g^2} ( G_{12} - S_{12} )\right)^{\text{dim}(U(N)/H)/2} \text{Vol}(U(N)/H) \prod_{ \alpha>0}\text{Tr}[E_{\alpha}, B]^2
\end{eqnarray}
\noindent
where the one-loop determinants, arising from the explicitly integration over the quantum fluctuation, have to be calculated. In this paper we don't perform in detail this calculation guessing that due
the supersymmetric properties of the background fields a cancelation between the bosonic and the fermionic contributions occurs. Indeed
this is what happens for the 1/2 BPS circular loop as shown in \cite{Okuda} and confirmed in \cite{PestunII}. The calculation for
a generic 't Hooft operators seems more intricate and thus, sketching the computation for the Zarembo 't Hooft operators in Appendix B, we leave the full analysis for a future investigation.\newline
In principle this is not the end of the story since another non-perturbative effect due to the so called monopole bubbling has to be considered. More in detail the presence of smooth monopole configurations that surround the singular monopole can screen the charge of the
't Hooft operator. Since the regular 't Hooft-Polyakov monopoles are labeled by the coroots, the net charge obtained from the screening
can be found through the action of the lower operator associated to the roots on the magnetic weight $B$. As consequence if the 't Hooft loop operator is in a certain representation all the weights that belong to it will contribute to the $VEV$ of the observable.

\subsection{Expectation value of DGRT-'t Hooft loops on $S^2$}
In this subsection we will compute explicitly the expectation value of the DGRT 't Hooft loops on $S^2$. Since we expect that their $VEV$ is a non trivial function of the coupling constant their analysis constitutes an intriguing playground to study the action of the $S$-duality.\newline
More in detail we consider the operators supported on a curve $C$ that lies on a two-dimensional sphere and parameterize by

\begin{equation}x_{\mu}(s) = ( x_1(s),  x_2(s), x_3(s), 0)\, \, \, \, \, \, \, 0 < s < 2 \pi\end{equation}
\noindent
with $x_{1}^2 + x_2^2 + x_3^2 =1$ and with the scalar couplings defined as in (\ref{DGRT})

\begin{equation}\theta^A(s) = M_K^A \epsilon^{IJK} \dot{x}_I(s)\, x_J(s).\end{equation}
\noindent
As noted in the previous section generically these observables are BPS and preserve four supersymmetries that are linear combinations of the super Poincar\'{e} and the super conformal charges. For the explicit calculation we take our operator in the fundamental representation identified by the magnetic weight

\begin{equation}B=\text{diag}\left( \underbrace{1,0\ldots 0,0}_{N} \right)\end{equation}
\noindent
in such a way that we also avoid the problem of the monopole bubbling phenomena\footnote{No screening of the 't Hooft operators charge can occur for the fundamental and for the k-antysimmetric representations of $U(N)$.}. The specified singular configuration breaks the $G = U(N)$ symmetry to $H = U(N-1) \times U(1)$, the coset space $G/H$ is isomorphic to $\mathbb{CP}^{N-1}$ and has dimension $2(N-1)$. Using that for a generic path on $S^2$
\begin{equation}G_{12} - S_{12} = 2 \frac{\mathcal{A}_1 \mathcal{A}_2}{\mathcal{A}^2}\end{equation}
\noindent
where $\mathcal{A}$ is the total area of the sphere and $\mathcal{A}_1$, $\mathcal{A}_2$ are the areas determined by the loop, the expectation value of the operators\footnote{We guess that the one-loop determinant factor is trivial as for the circular magnetic operators. A proof of this fact is under investigation.} at the leading order is given by

\begin{equation}\langle H_F \rangle = e^{-\frac{8 \pi^2}{g^2}\frac{\mathcal{A}_1 \mathcal{A}_2}{\mathcal{A}^2}} \left(\frac{16 \pi^2}{g^2}\frac{\mathcal{A}_1 \mathcal{A}_2}{\mathcal{A}^2}\right)^{N-1} \frac{1}{(N-1)!}\, \, .\label{44}\end{equation}
\noindent
This formula is valid at the leading order in the strong coupling expansion and can be interestingly compared with the expectation value of the $S$-dual electric operator in order to understand how the $S$-duality acts on such class of observables. We recall that for the DGRT Wilson loops \cite{Pestun,Drukker:2007yx,Drukker:2007dw,Nostro,Young} a conjecture relates their expectation value to the $VEV$ of the ordinary Wilson loops in the zero-instanton sector of the pure bosonic two-dimensional Yang-Mills theory on $S^2$. Since in that case the theory is completely solvable \cite{Migdal:1975zg}, an exact expression for their $VEV$ can be derived \cite{Kazakov:1980zi}\cite{Rusakov:1990rs}\cite{Bassetto:1998sr}

\begin{equation}\langle W_F \rangle = \frac{1}{N} L_{N-1}^1 \left(g_{2d}^2 \frac{\mathcal{A}_1 \mathcal{A}_2}{\mathcal{A}}\right) \text{exp}\left( -\frac{g_{2d}^2}{2} \frac{\mathcal{A}_1 \mathcal{A}_2}{\mathcal{A}}\right)\end{equation}
\noindent
where $g^2_{2d}$ is the two-dimensional coupling constant. More precisely the conjecture states that the $VEV$ of the DGRT-Wilson loop in ${\cal N}$ = 4 SYM is obtained from the previous formula trough a redefinition of the two dimensional coupling constant

\begin{equation}g^2_{2d} = \frac{g^2}{\mathcal{A}}.\end{equation}
\noindent
Now performing a strong coupling expansion of the result, since the Laguerre polynomial in this regime is replaced by its argument to the
 maximal power, one obtains that

\begin{equation}\langle W_F \rangle \simeq \frac{1}{(N-1)!}\left(g^2 \frac{\mathcal{A}_1 \mathcal{A}_2}{\mathcal{A}^2}\right)^{N-1} \text{exp}\left( -\frac{g^2}{2} \frac{\mathcal{A}_1 \mathcal{A}_2}{\mathcal{A}^2}\right).\end{equation}
\noindent
Remarkably the previous expression is identical to the formula (\ref{44}) for the expectation value of the 't Hooft operator after the usual $S$-dual transformations on the coupling constant

\begin{equation}g^2 \rightarrow g'^{\, 2} = \frac{16 \pi^2}{g^2}.\end{equation}
\noindent
Indeed since $S$-duality is a generalization of the ordinary electric-magnetic duality, it maps Wilson operators in a theory with
coupling constant $g$ to 't Hooft operators, defined on the same circuit and with the same scalar couplings,  in a theory with coupling constant $g'$\footnote{Since we are dealing with the $ G= ^LG = U(N)$ gauge group the representation of the group and its dual are in one-to-one correspondence}.

\section{Summary and discussions}
\normalsize
In this paper we have defined a large family of BPS 't Hooft loop operators in ${\cal N}$= 4 SYM theory. Indeed, even if
the $S$-duality predicts their existence, no magnetic operators preserving less supersymmetries than the line and the circular 't Hooft loops has been analyzed in the literature.\newline
Starting from the ${\cal N}$= 4 Maxwell theory, in section two we have introduced the mixed BPS Wilson-'t Hooft operators : after their definition we focused on their supersymmetric properties deriving the BPS condition for a generic line operator and on the computation of their expectation value.\newline
In the following step the generalization of these observables to the non-abelian $U(N)$ ${\cal N}$= 4 SYM theory has been discussed. More in detail the magnetic configurations associated to Zarembo and DGRT 't Hooft loops have been explicitly shown and their supersymmetric properties have been carefully studied. Interestingly we have found that the supercharges preserved by the BPS Wilson loops and by their magnetic counterparts are not the same but are related by a four-dimensional chiral transformation as predicted from $S$-duality. Furthermore the quantum definition of the operators is discussed and the expectation value for a specific class of DGRT-'t Hooft loops has been calculated up to next-to-leading order in perturbation theory.\newline
More work has to been done to have a complete picture of these operators. The first interesting problem to investigate regards the analysis of the dyonic loop observables in ${\cal N}$= 4 SYM. Their construction, presented in section two only for the Maxwell theory, in ${\cal N}$= 4 SYM seems to be more intricate respect to the definition of the pure magnetic operators. Indeed in order to describe them one has not only to impose the boundary conditions for the fields as in the magnetic case but also to introduce in the path integral a Wilson operator in a certain irreducible representation $R$ of the stabilizer subgroup of the magnetic weight $B$ \cite{Kapustin:2005py} and the analysis of the supersymmetric properties can not be straighforward derived from the abelian case.\newline
To complete the next-to-leading order calculation of the $VEV$ of these operators performed in section (4), the evaluation of the one-loop determinants around the singular configurations is necessary. Perform explicitly the computation by diagonalizing the fluctuations operators seems quite intricate and indeed in \cite{Gomis:2011pf} the relevant determinants have been calculated using the Atiyah-Singer index theorem. It would be intriguing to extend this computation for at least some classes of magnetic operators defined in this paper.\newline
Different directions can be investigate as extensions of the present work. One can start from the study of the correlator of 't Hooft loops whose singular configurations associated should be a simple sum of that generated by the two 't Hooft loops on the two different paths. Furthermore in order to analyze the Operator Product Expansion (OPE) of the 't Hooft operators and extract additional information about the action of the $S$-duality one could investigate also their correlator with some local objects as the chiral primary operators (see \cite{Gomis:2009xg} for the correlators between 1/2 BPS magnetic osbervable and CPOs).\newline
Another interesting direction for a future investigation is given by the definition and the study of these magnetic and mixed operators in some theories like the ${\cal N}$= 2 SYM where, differently from the ${\cal N}$ = 4 SYM, the action of the $S$-duality is highly non trivial \cite{poss}\cite{Drukker:2009id}\cite{Drukker:2010jp}. The generalization of the construction to some three dimensional theory would be equally interesting in order to investigate the properties of the 3d mirror symmetry \cite{Borokhov:2002ib}\cite{Borokhov:2002cg}.\newline
Finally we could go further in the analysis on the relation between the DGRT loops operators in ${\cal N}$ = 4 SYM and the two dimensional Yang-Mills theory. In \cite{PestunII} the authors conjectured that the $VEV$ of the 1/2 BPS circular 't Hooft operators is equivalent to the value of the two dimensional Yang-Mills partition function on the two sphere in a non-zero instanton sector. The situation is not so clear if we instead of the maximal circle we choose a generic path on $S^2$. Thus It could be intriguing first to prove rigourously the conjecture and then identifies, if they exist, the correspondent two-dimensional observables of the DGRT 't Hooft loops.

\newpage

\section*{Acknowledgments}

I am especially grateful to Domenico Seminara and Luca Griguolo for valuable discussion at different stages of this work, for reading of the draft and together with Gabriele Martelloni for participating at the early stage of
the project. It's a pleasure to thank for the
hospitality the G. Galilei Institute for Theoretical
Physics in Florence and the theory group
at the University of Turin where this work started.
I am supported by the Research Executive Agency (REA)
of the European Union under Grant Agreement PITNGA-
2009-238353 (ITN STRONGnet).

\appendix
\section{Conventions}
The four dimensional ${\cal N}$ = 4 SYM can be obtained from the dimensional reduction of the
${\cal N}$ = 1 SYM in $d=10$ and its action can be written using the notation of \cite{Pestun:2007rz} as

\begin{equation}S = \frac{1}{g^2}\int d^4x \frac{1}{2} F_{MN} F^{MN} -\, \Psi \gamma^M D_M \Psi \end{equation}
\noindent
where all the field ($A_M,\psi)$ are in the adjoint representation of the gauge group $G$ and $M$, $N$ take values in $\{0..9\}$.
The covariant derivative and the field strength are given respectively by $D_M= \partial_M + A_M$ and $F_{MN} = [ D_M, D_N ]$.
The space-time indices running from 1 to 4 have been indicated by Greek letters $\mu, \nu, \rho, \sigma...$ while the six directions associated with the R-symmetry have been labeled by the letter $A,B...$ with values in $\{0,5,6,7,8,9\}$.\newline
The $32 \times 32$ gamma matrices $\gamma^M$ satisfy the anti-commutation rules

\begin{equation}\{\gamma^M, \gamma^N \} = 2 \delta^{MN}\end{equation}
\noindent
and in the Weyl representation can be taken in the following form

\begin{center}
$\gamma^M =
\left(
  \begin{array}{cc}
  0 & \tilde{\Gamma}^M \\
  \Gamma^M & 0 \\
  \end{array}
\right)$
\end{center}
\noindent
where the explicitly expression of the $16 \times 16$ matrices $\Gamma^M = ( \tilde{\Gamma}^M )^{\dagger}$ can be found in the Appendix A of \cite{Pestun:2007rz}. In this representation the
 Dirac spinor $\psi$ splits into two sixteen component spinors of opposite chirality $\psi=(\psi_+, \psi_-)$ (respect to the chiral matrix $\gamma^{11}=-i \gamma^1...\gamma^9 \gamma^{0}$).\newline
The ${\cal N}$ = 4 SYM action is left invariant by the super-conformal transformations

\begin{equation}\delta_{\epsilon} A_M(x) = \epsilon(x) \Gamma^M \psi\end{equation}

\begin{equation}\delta_{\epsilon} \psi = \frac{1}{2} F_{MN} \Gamma^{MN} \epsilon(x) - 2 \tilde{\Gamma }^A \phi^A \epsilon_1\end{equation}
\noindent
where the spinor $\epsilon(x)$ is a conformal Killing spinor on $\mathbb{R}^4$

\begin{equation}\epsilon(x) = \epsilon_0 + \epsilon_{1} x_{\mu} \Gamma^{\mu},\end{equation}
\noindent
 with $\epsilon_0$ and $\epsilon_1$ two $16$ component constant spinors that generate the usual Super Poincar\'{e} and Super Conformal symmetries respectively.

\section{One-loop determinant around the singular configurations}
The computation of the one loop determinants for the 1/2 BPS 't Hooft operator has been performed in \cite{Okuda}
by diagonalizing the quadratic fluctuation operator around the singular configurations and more recently in \cite{Gomis:2011pf}
where instead the Atiyah-Singer index theorem has been used to perform the calculation. In this appendix following the approach of
\cite{Okuda} we try to extend the computation to the BPS magnetic operators introduced in this paper.\newline
The one-loop determinant factor, obtained integrating out the quantum fluctuations, can be written in a compact ten-dimensional notation as

\begin{equation}\frac{[ \text{det}( i \Gamma^M D_{0\,M})]^{1/4}\, \,\text{det}_g(-D_0^2)}{[\text{det}_b(-\delta^{MN}D_0^2 + 2 i F_0^{MN})]^{1/2}}\label{RT}\end{equation}
\noindent
where the covariant derivative $D_{0\, M}$ is in the background of the classical fields $A_{\mu}^0, \phi_A^0$. In the paper we guess that a cancelation between the fermionic and bosonic determinants, that make the expression (\ref{RT}) trivial, occurs not only for the magnetic line operator but for every BPS configurations. The physical reason can be explained as follow. If we suppose to have an eigenfunction of the scalar operator a supersymmetric transformation that leaves the background invariant should rotate it into an eigenfunction of the fermionic operator and thus relate their corresponding eigenvalue. Since an analogous relation holds also between the gluino and the gluon spectra, rewriting (\ref{RT}) as a product over the eigenvalues and taking carefully their multiplicity, one could in principle see that the cancelation around the BPS background is a general fact of the supersymmetric theories \cite{D'Adda:1977ur}. Let us sketch briefly the idea for the Zarembo 't Hooft loops leaving the complete proof and the analysis of more complex DGRT 't Hooft operators for a future investigation.\newline
We start by supposing to know the eigenfunctions of the bosonic operators $A^{\lambda}_{N}$\footnote{Here all the indices A, B, M, N run from 0 to 9} with eigenvalue $\lambda^2$
\noindent
\begin{equation}\left(- \delta^{MN} D^2 + 2 i F^{MN} \right) A^{\lambda}_{N} = \lambda^2  A^{\lambda}_{M}.\label{vv}\end{equation}
\noindent
From them one can construct the functions $\Psi_{\lambda}$ as

\begin{equation}\Psi^{\lambda} = \Gamma^{AM} D_A^0 A^{\lambda}_M \epsilon_0\end{equation}
\noindent
where the spinor $\epsilon_0$ is a solution of the BPS equations for the Zarembo 't Hooft operator (see section 3.3), \emph{i.e.}

\begin{equation}\frac{1}{2} \Gamma^{AB} F^0_{AB} \epsilon_0 =0\, .\end{equation}
\noindent
It's not so difficult to see that $\Psi_{\lambda}$ defined in such way are eigenfunctions of the fermionic operator with eigenvalue $\lambda$ since
\begin{equation}\left( i \Gamma^M D^0_M \left( i \Gamma^N D^0_N \Psi_{\lambda}\right)\right) = \left( i \Gamma^M D^0_M \left( - i \lambda^2 \Gamma^N A_N^{\lambda} \epsilon_0 \right)\right) = \lambda^2 \Gamma^{AM} D_A^0 A^{\lambda}_M \epsilon_0  = \lambda^2 \Psi^{\lambda}\label{vv}\end{equation}
\noindent
Furthermore choosing the autofunction of the scalar operator as

\begin{equation}\varphi^{\lambda} = \epsilon_0 \Gamma^{AM} D_{0 A} A_M \epsilon_0\end{equation}
\noindent
we have that

\begin{equation}- D_0^2 \varphi^{\lambda} = - \epsilon_0 D_0^2 (\Gamma^{AM} D_{0 A} A_M )\epsilon_0 = \epsilon_0 \lambda^2 \Gamma^{AM} D_{0 A} A_M \epsilon_0 = \lambda^2 \varphi^{\lambda} \label{vvv}\end{equation}
\noindent
Now rewriting the determinants as a product over the eigenvalues and taking carefully their correct multiplicity it's simple to see from
(\ref{vv}) and (\ref{vvv}) that the computation simplifies drastically giving the trivial result.\newline
This conclude the calculation for the Zarembo 't Hooft loop but unfortunately this construction can not be straightforward generalized to the DGRT 't Hooft loops or in general to BPS operators preserving some super-conformal charges. Probably the fact that conformal symmetries is explicitly broken by the gauge fixing term makes the computation more intricate respect to the one presented here.

\section{Integration over the Adjoint Orbits}

In this appendix we review some elements that we have used in the integration over the adjoint orbit of the coweight $B$. Let us consider the
Lie algebra $g$ associated to the group $G$. A common notation is to indicate with $H_i$ the generator of the Cartan subalgebra of $g$ and with $E_{\pm \alpha}$ the ladder operators associated with the root $\alpha$.\newline
Now let $B = b_i H_i$ a magnetic weight (coweight) with value in the Cartan algebra $h$, its adjoint orbit $O_B$ is defined as

\begin{equation}O_B \equiv \{ B_g = g B g^{-1}\, , g \text{ in } G\}\end{equation}
\noindent
and it is diffeomorphic to the coset space $G/H$ where $H$ is the invariance group of $B$. Following \cite{Okuda} one can construct the metric
on $O_B$ from the Maurer-Cartan one-form

\begin{equation} g^{-1}dg = i \left( \sum_i d\xi^i H_i + \sum_{\alpha} d\xi^{\alpha} E_{\alpha}\right)\end{equation}
\noindent
and find that

\begin{equation}ds^2_{O_B} = 2 \mathcal{N} \sum_{a>0} \alpha(B)^2 \text{Tr}( E_{\alpha}, E_{-\alpha})|d\xi^{\alpha}|^2 \end{equation}
\noindent
where $\mathcal{N}$ is a normalization factor fixed from the value of the on-shell action and the sum is done over the positive root elements that don't belong to the invariance group of B, namely all $E_{\alpha}$ that satisfy $ [ E_{\alpha}, B ] = \alpha(B) E_{\alpha} \neq 0$. The integration over the orbit give thus

\begin{equation}\int ds^2_{O_B} = \left(\frac{\mathcal{N}}{\pi}\right)^{\text{dim(G/H)/2}} \text{Vol}(G/H) \prod_{\alpha(B)\neq 0,\, \alpha>0} \alpha(B)^2.\end{equation}

\noindent
In the paper we have used the formula for the volume of the compact group $U(N)$ \cite{mcdonald}

\begin{equation}\text{Vol}(U(N)) = \frac{(2\pi)^{N(N+1)/2}}{\prod_{n=1}^{N-1}n!}\end{equation}

\section{'t Hooft loop configurations}\label{wedge}
\subsection{Zarembo - 't Hooft loop operators : 1/4 BPS cusp }
In the following we present an explicit configuration for 1/4 BPS Zarembo 't Hooft loop, \emph{i.e.} a magnetic operator supported on a cusp at angle $\alpha$. The circuit can be parameterized as

\small
\begin{equation}x_{\mu}(s) = \{ s, 0, 0, 0 \}\, \, \, \, \text{for}\, \,-\infty < s < 0, \hspace{1cm} x_{\mu}(s) = \{ - s\, \cos{\alpha}, - s\, \sin{\alpha}, 0, 0 \}\, \, \, \text{for}\, \, 0 < s < + \infty\end{equation}
\normalsize
\noindent
while the scalar coupling and the matrix $M_{\mu}^A$ can be chosen without loss of generality as
\begin{center}$\theta^A(s)\, = M_{\mu}^A \dot{x}^{\mu}(s)$ \label{M}\hspace{1cm}
$M_{\mu}^{A}$ = $\left(
  \begin{array}{cccccc}
    0 & 0 & 0 & 0 & 0 & -1 \\
    0 & 0 & 0 & 0 & -1 & 0 \\
    0 & 0 & 0 & -1 & 0 & 0 \\
    0 & 0 & -1 & 0 & 0 & 0 \\
  \end{array}
\right).$
\end{center}
\noindent
The classical configurations associated to the operator are given by
\small
\begin{equation}A_{1}(y) = B \frac{\pi ( 1 - \cos{\alpha} )  - 2 \text{ArcTan}\left[\frac{y_1}{y_{\bot_1}^2}\right] ( 1 + \cos{\alpha} )}{2\, y_{\bot_1}^2}\, \, \, \, \, \, \, \, \, \, A_{2}(y) = - B \sin{\alpha} \frac{\pi - 2 \text{ArcTan}\left[\frac{y_2}{y_{\bot_2}^2}\right]}{2\, y_{\bot_2}^2}\end{equation}

\begin{equation}\phi^9(y) = - B \frac{\pi ( 1 - \cos{\alpha} ) - 2  \text{ArcTan}\left[\frac{y_1}{y_{\bot_1}^2}\right]( 1 + \cos{\alpha} )}{2\, y_{\bot_1}^2}\, \, \, \, \, \,\phi^8(y) =  B \sin{\alpha} \frac{\pi - 2 \text{ArcTan}\left[\frac{y_2}{y_{\bot_2}^2}\right]}{2\, y_{\bot_2}^2}\end{equation}
\normalsize
\noindent
where $y_{\bot_1}^2= y_2^2+y_3^2+y_4^2$ and $y_{\bot_2}^2= y_1^2+y_3^2+y_4^2$. Since the supersymmetric variation of the operator is given by two independent relations for the spinor $\epsilon_0$

\begin{equation}\left(  \Gamma^{234} +  \Gamma^{9} \right)\epsilon_0 =0\, \, \, \, \, \, \, \, \, \,  \left(  \Gamma^{134} -  \Gamma^8 \right)\epsilon_0 =0\end{equation}
\noindent
as expected the magnetic "cusp" operator is a 1/4 BPS object.

\subsection{DGRT - 't Hooft loop operators : 1/4 BPS latitude}

\normalsize
\noindent
For the DGRT magnetic operator supported on a latitude at polar angle $\theta_0$ we parameterize the circuit as

\begin{equation}x_{\mu}(s)= ( \sin{\theta_0}\, \cos{s}, \sin{\theta_0} \sin{s}, \cos{\theta_0},0)\end{equation}
\noindent
with $0 < s < 2 \pi$ and choose the matrix $M$  that defines the scalar couplings (\ref{DGRT}) of the form

\begin{center}\label{M2}
$M_{K}^{A}$ = $\left(
  \begin{array}{cccccc}
    0 & 0 & 0 & 1 & 0 & 0 \\
    0 & 0 & 0 & 0 & 1 & 0 \\
    0 & 0 & 0 & 0 & 0 & 1 \\
  \end{array}
\right)\label{Podd}$
\end{center}
\noindent
even if other choices of the matrix are allowed due the invariance of the theory under the super-conformal group. As in the electric case the new 't Hooft operator, differently from the maximal circle one, couples with three of the six scalars of the theory. Their classical configurations have these behaviors

\small
\begin{equation}\phi^7(y) = - 2 \pi B \frac{y_1 \cos (\theta_0 ) \left( 1 + y^2 -2 y_3 \cos (\theta_0
   )- y_c \right)}{y_{\|}^2\, \, y_c}\end{equation}
\noindent
\begin{equation}\phi^8(y) =  2 \pi B  \frac{y_2 \cos (\theta_0 ) \left( 1 + y^2 -2 y_3 \cos (\theta_0
   )- y_c \right)}{ y_{\|}^2\, \, y_c}\, \, \, \, \, \, \, \, \, \phi^9(y) = 2 \pi B \frac{\sin{(\theta_0)}^2}{y_c}\end{equation}
\normalsize
\noindent
while for the gauge fields we have that

\small
\begin{equation}A_1(y) = - 2 \pi B  \frac{y_2 \left( 1 + y^2 -2 y_3 \cos (\theta_0
   )- y_c \right)}{2\,  y_{\|}^2\, \, y_c}\, \, \, \, \, \, \, \, \,  A_2(y) = 2 \pi B \frac{y_1 \left( 1 + y^2 -2 y_3 \cos (\theta_0
   )- y_c \right)}{ y_{\|}^2\, \, y_c} \end{equation}
\noindent
\normalsize
where in order to make more compact the expressions we made use of the notation

\small
\begin{displaymath}y_c =  \sqrt{(1+y^2 - 2 y_3 \cos (\theta_0))^2 - 4 y_{\|}^2 \sin (\theta_0)^2},\, \, \, \, \, y^2 = y_1^2+y_2^2+y_3^2+y_4^2,\, \, \, \,  y_{\|}^2 = y_1^2+y_2^2.\end{displaymath}
\normalsize
From these expressions one can immediately check that in the case in which the latitude chosen is the equator they reduces to the well known 1/2 BPS circular configurations shown in the previous subsection.\newline
The supersymmetric variation of the operator reduces to two independent relations for the spinor $\epsilon_0$ and $\epsilon_1$ :

\begin{equation}( \Gamma^{82} + \Gamma^{71} ) \epsilon_1 = 0 \label{T}\end{equation}

\begin{equation} \Gamma^{9} \epsilon_0 =  (- \Gamma^{34} - ( \Gamma^{82} + \Gamma^{93} ) \cos{\theta_0}) \epsilon_1 \label{TT}\end{equation}
\noindent
Again these equations become exactly those that describe the 1/4 BPS circular Wilson loops \cite{Drukker} at polar angle $\theta_0$

\begin{equation}( \Gamma^{82} + \Gamma^{71} ) \epsilon_1 = 0\end{equation}

\begin{equation} \Gamma^{9} \epsilon_0 =  ( i \Gamma^{12} - ( \Gamma^{82} + \Gamma^{93} ) \cos{\theta_0}) \epsilon_1 \end{equation}
\noindent
after the usual chiral transformation on the spinors (\ref{BO}).

\subsection{DGRT - 't Hooft loop operators : 1/4 BPS wedge}\label{wedge}

In this sub-appendix we present the singular configurations associated to the 1/4 BPS magnetic operator supported on a "wedge" namely a circuit consisting of two longitudes separated by an azimuthal angle $\delta$ on $S^2$.  The circuit is specified by

\begin{equation}x_{\mu}(s)= ( \sin{(s)} , 0, \cos{(s)}, 0 )\, \, \hspace{1cm} 0<s<\pi\end{equation}

\begin{equation}x_{\mu}(s)= ( -\cos{\delta}\sin{(s)} , -\sin{\delta}\sin{(s)}, \cos{(s)}, 0 )\, \, \hspace{1cm} \pi<s<2\pi\end{equation}
\noindent
and the scalar couplings are given exactly as in the previous example by

\begin{equation}\theta^A(s)\, = \epsilon_{IJK} M_{I}^A \dot{x}_{J}(s) x_K(s) \end{equation}
\noindent
with the matrix $M$ shown in (\ref{Podd}). The configurations of the gauge and the scalar fields are given by

\small
\begin{displaymath}A_1 =\frac{B}{2}\frac{2 y_3 ( 1 + y^2 ) ( \pi + 2 \text{ArcTan}(\frac{2 y_1}{y_c}) - 2 y_c ( \pi y_3 + y_1 \text{Log}[1 + \frac{4 y_3}{1 + y^2 - 2 y_3}])}{ (y_{1}^2 + y_3^2) y_c} - \end{displaymath}
\begin{equation}- \frac{1}{4} \cos{\delta}\frac{2 y_3 ( 1 + y^2 ) ( \pi + 2 \text{ArcTan}(\frac{2 y_t}{y_s}) - 2 y_s( \pi y_3 + y_t \text{Log}[1 + \frac{4 y_3}{1 + y^2 - 2 y_3}])}{(y_t^2 + y_3^2) y_s} \end{equation}

\begin{equation}A_2 = - \frac{B}{2} \sin{\delta}\frac{2 y_3 ( 1 + y^2 ) ( \pi + 2 \text{ArcTan}(\frac{2 y_t}{y_s}) - 2 y_s( \pi y_3 + y_t \text{Log}[1 + \frac{4 y_3}{1 + y^2 - 2 y_3}])}{(y_t^2 + y_3^2) y_s} \end{equation}

\begin{displaymath}A_3 =\frac{B}{2}\frac{2 y_3 ( 1 + y^2 ) ( \pi + 2 \text{ArcTan}(\frac{2 y_1}{y_c}) - 2 y_c ( \pi y_3 + y_1 \text{Log}[1 + \frac{4 y_3}{1 + y^2 - 2 y_3}])}{ (y_{1}^2 + y_3^2) y_c} - \end{displaymath}
\begin{equation}- \frac{1}{4} \cos{\delta}\frac{2 y_3 ( 1 + y^2 ) ( \pi + 2 \text{ArcTan}(\frac{2 y_t}{y_s}) - 2 y_s( \pi y_3 + y_t \text{Log}[1 + \frac{4 y_3}{1 + y^2 - 2 y_3}])}{(y_t^2 + y_3^2) y_s} \end{equation}

\begin{equation}A_4 = 0\end{equation}

\begin{equation}\phi^8 =  \frac{B}{2}\left(\frac{( \pi + 2 \text{ArcTan}(\frac{2 y_0}{y_c}))}{  y_c} - \cos{\delta}\frac{( \pi + 2 \text{ArcTan}(\frac{2 y_t}{y_s}))}{  y_s}\right) \end{equation}

\begin{equation}\phi^7 = \frac{B}{2}\left( \sin{\delta}\frac{( \pi + 2 \text{ArcTan}(\frac{2 y_t}{y_s}))}{  y_s}\right) \, \, \, \hspace{1cm} \phi_i = 0\, \, \,\hspace{0.2cm} \text{for } i=4,5,6,9 \end{equation}
\noindent
\normalsize
where we have used for convenience the following notation
\small
\begin{displaymath}y_c =  \sqrt{1+y^2 - 4 y_{\|}^2}\, \, \hspace{1.3cm} y_s =  \sqrt{1+y^2 - 4 y_{3}^2 - 4 ( y_1 \cos{\delta} + y_2 \sin{\delta} )^2 }\end{displaymath}
 \begin{displaymath}y_t =  ( y_1 \cos{\delta} + y_2 \sin{\delta} )\, \, \hspace{1cm} y^2 = y_1^2+y_2^2+y_3^2+y_4^2\, \, \hspace{0.3cm} y_{\|}^2 = y_1^2+y_3^2.\end{displaymath}\noindent
\normalsize
As a first simple check is immediate to see that for $\delta=\pi$ these configurations reduce to the maximal circle ones (see section \ref{maxim}). Calculating the supersymmetric variation of the configurations we obtain  two independent equations

\begin{equation}  \Gamma^{24} \, \epsilon_1+ \Gamma^{8}\,  \epsilon_0 = 0\end{equation}

\begin{equation}[ ( \Gamma^{24} \cos{\delta} - \Gamma^{14} \sin{\delta})] \epsilon_1 +  ( \Gamma^{8} \cos{\delta} - \Gamma^{7}\sin{\delta}) \epsilon_0 = 0\end{equation}
\noindent
and as consequence the operator is 1/4 BPS. One can compare in detail these supercharges with those preserved
by the dual electric operator and that are given by the solutions of

\begin{equation} i \Gamma^{13} \, \epsilon_1+ \Gamma^{8}\,  \epsilon_0 = 0\end{equation}

\begin{equation}[ i (\Gamma^{13} \cos{\delta} + \Gamma^{23} \sin{\delta})] \epsilon_1 +  ( \Gamma^{8} \cos{\delta} - \Gamma^{7}\sin{\delta}) \epsilon_0 = 0\end{equation}
\noindent
to verify again that the $S$-duality acts on them as expected.


\begin{thebibliography}{99}

\bibitem{Pestun:2007rz}
  V.~Pestun,
  ''Localization of gauge theory on a four-sphere and supersymmetric Wilson
  loops,''
  arXiv:0712.2824 [hep-th]

\bibitem{Gomis:2011pf}
  J.~Gomis, T.~Okuda and V.~Pestun,
  ``Exact Results for 't Hooft Loops in Gauge Theories on $S^4$,''
  arXiv:1105.2568 [hep-th]

\bibitem{TAKI}
  Y.~Ito, T.~Okuda, M.~Taki,
  ``Line operators on $S^1 \times \mathbb{R}^3$ and quantization of the Hitchin moduli space,''
  arXiv:0712.2824 [hep-th]



 \bibitem{Rey:1998ik}
  S.~-J.~Rey and J.~-T.~Yee,
  ``Macroscopic strings as heavy quarks in large N gauge theory and anti-de Sitter supergravity,''
  Eur.\ Phys.\ J.\ C {\bf 22} (2001) 379
  [hep-th/9803001].

\bibitem{Maldacena:1998im}
  J.~M.~Maldacena,
 ``Wilson loops in large N field theories,''
 Phys.\ Rev.\ Lett.\  {\bf 80} (1998) 4859
 [arXiv:hep-th/9803002]

\bibitem{Zarembo:2002an}
  K.~Zarembo, "Supersymmetric Wilson loops,"
  Nucl.\ Phys.\  B {\bf 643}, 157 (2002)
  [arXiv:~hep-th/0205160].

\bibitem{Drukker:2000rr}
  N.~Drukker and D.~J.~Gross,
  ``An Exact prediction of N=4 SUSYM theory for string theory,''
  J.\ Math.\ Phys.\  {\bf 42}, 2896 (2001)
  [arXiv:hep-th/0010274].

\bibitem{Er} J.K. Erickson, G.W. Semenoff, K. Zarembo,
{ Wilson loops in N=4 supersymmetric Yang-Mills theory},
Nucl.\ Phys.\ B {\bf 582} (2000) 155
[hep-th/0003055].










\bibitem{Drukker:2007dw}
N.~Drukker, S.~Giombi, R.~Ricci and D.~Trancanelli,
"More supersymmetric Wilson loops,"
Phys.\ Rev.\  D {\bf 76} (2007) 107703
 [arXiv:0704.2237 [hep-th]].


\bibitem{Drukker:2007yx}
N.~Drukker, S.~Giombi, R.~Ricci and D.~Trancanelli,
"Wilson loops: From four-dimensional SYM to two-dimensional YM,"
 Phys.\ Rev.\  D {\bf 77} (2008) 047901
  [arXiv:0707.2699 [hep-th]].

\bibitem{Drukker:2007qr}
  N.~Drukker, S.~Giombi, R.~Ricci and D.~Trancanelli,
"Supersymmetric Wilson loops on $S^3$,"
  [arXiv:~hep-th/0711.3226].


\bibitem{Pestun}
  V.~Pestun,
  "Localization of the four-dimensional N=4 SYM to a two-sphere and 1/8 BPS
  Wilson loops,"
  arXiv:0906.0638 [hep-th].

\bibitem{PestunII}
  S.~Giombi and V.~Pestun,
  "Correlators of local operators and 1/8 BPS Wilson loops on $S^2$ from 2d YM
  and matrix models,"
  JHEP {\bf 1010} (2010) 033
  [arXiv:0906.1572 [hep-th]].

\bibitem{Nostro}
  A.~Bassetto, L.~Griguolo, F.~Pucci and D.~Seminara,
  "Supersymmetric Wilson loops at two loops,"
  JHEP {\bf 0806} (2008) 083
  [arXiv:0804.3973 [hep-th]].

\bibitem{Young}
  D.~Young,
  "BPS Wilson Loops on $S^2$ at Higher Loops,"
  JHEP {\bf 0805} (2008) 077
  [arXiv:0804.4098 [hep-th]].

\bibitem{CorrelatoriI}
  A.~Bassetto, L.~Griguolo, F.~Pucci, D.~Seminara, S.~Thambyahpillai and D.~Young,
  "Correlators of supersymmetric Wilson-loops, protected operators and matrix
  models in N=4 SYM,"
  JHEP {\bf 0908} (2009) 061
  [arXiv:0905.1943 [hep-th]].


  \bibitem{CorrelatoriII}
  A.~Bassetto, L.~Griguolo, F.~Pucci, D.~Seminara, S.~Thambyahpillai and D.~Young,
  "Correlators of supersymmetric Wilson loops at weak and strong coupling,"
  JHEP {\bf 1003} (2010) 038
  [arXiv:0912.5440 [hep-th]].



\bibitem{Dymarsky:2009si}
  A.~Dymarsky and V.~Pestun,
  "Supersymmetric Wilson loops in N=4 SYM and pure spinors,"
  JHEP {\bf 1004} (2010) 115
  [arXiv:0911.1841 [hep-th]].

\bibitem{thooft}
  G.~'t Hooft,
  "On the Phase Transition Towards Permanent Quark Confinement,"
  Nucl.\ Phys.\ B {\bf 138} (1978) 1.

  \bibitem{dyn}
  E.~Witten,
  "Dyons Of Charge E Theta/2 Pi,"
  Phys.\ Lett.\  B {\bf 86} (1979) 283.

  \bibitem{GNO}
  P.~Goddard, J.~Nuyts and D.~I.~Olive,
  "Gauge Theories And Magnetic Charge,"
  Nucl.\ Phys.\  B {\bf 125} (1977) 1.



\bibitem{Okuda}
  J.~Gomis, T.~Okuda and D.~Trancanelli,
  "Quantum 't Hooft operators and S-duality in N=4 super Yang-Mills,"
  Adv.\ Theor.\ Math.\ Phys.\  {\bf 13} (2009) 1941
  [arXiv:0904.4486 [hep-th]].

\bibitem{Gomis:2009xg}
  J.~Gomis and T.~Okuda,
  "S-duality, 't Hooft operators and the operator product expansion,"
  JHEP {\bf 0909} (2009) 072
  [arXiv:0906.3011 [hep-th]].

\bibitem{Montonen:1977sn}
  C.~Montonen and D.~I.~Olive,
 ``Magnetic Monopoles As Gauge Particles?,''
  Phys.\ Lett.\  B {\bf 72} (1977) 117.


\bibitem{WittenOlive}
  E.~Witten and D.~I.~Olive,
 ``Supersymmetry Algebras That Include Topological Charges,''
  Phys.\ Lett.\  B {\bf 78} (1978) 97.

\bibitem{Osborn:1979tq}
  H.~Osborn,
  "Topological Charges For N=4 Supersymmetric Gauge Theories And Monopoles Of

\bibitem{A}
  K.~A.~Intriligator,
  "Bonus symmetries of N=4 superYang-Mills correlation functions via AdS duality,"
  Nucl.\ Phys.\ B {\bf 551} (1999) 575
  [hep-th/9811047].

\bibitem{B}
  K.~A.~Intriligator and W.~Skiba,
  "Bonus symmetry and the operator product expansion of ${\cal N}=4$ SuperYang-Mills,"
  Nucl.\ Phys.\ B {\bf 559}, 165 (1999)
  [hep-th/9905020].

  \bibitem{C}
  P.~C.~Argyres, A.~Kapustin and N.~Seiberg,
  "On S-duality for non-simply-laced gauge groups,"
  JHEP {\bf 0606} (2006) 043
  [hep-th/0603048].

\bibitem{D}
  S.~Gukov and E.~Witten,
  "Gauge Theory, Ramification, And The Geometric Langlands Program,"
  hep-th/0612073.

  \bibitem{E}
  J.~Gomis and S.~Matsuura,
  "Bubbling surface operators and S-duality,"
  JHEP {\bf 0706} (2007) 025
  [arXiv:0704.1657 [hep-th]].

  \bibitem{F}
  D.~Gaiotto and E.~Witten,
  "Supersymmetric Boundary Conditions in N=4 Super Yang-Mills Theory,"
  arXiv:0804.2902 [hep-th].

  \bibitem{G}
  D.~Gaiotto and E.~Witten,
  "S-Duality of Boundary Conditions In N=4 Super Yang-Mills Theory,"
  arXiv:0807.3720 [hep-th].



  Spin 1,"  Phys.\ Lett.\  B {\bf 83} (1979) 321.
\bibitem{Kapustin:2005py}
  A.~Kapustin,
  "Wilson-'t Hooft operators in four-dimensional gauge theories and
  $S$-duality,"
  Phys.\ Rev.\  D {\bf 74} (2006) 025005
  [arXiv:hep-th/0501015]


\bibitem{Kapustin:2006pk}
  A.~Kapustin and E.~Witten,
  "`Electric-magnetic duality and the geometric Langlands program,"
  arXiv:hep-th/0604151.

\bibitem{Saulina:2011qr}
  N.~Saulina,
  "A note on Wilson-'t Hooft operators,"
  Nucl.\ Phys.\  B {\bf 857} (2012) 153
  [arXiv:1110.3354 [hep-th]].

\bibitem{Moraru:2012nu}
  R.~Moraru and N.~Saulina,
  "OPE of Wilson-'t Hooft operators in N=4 and N=2 SYM with gauge group G=PSU(3),"
  arXiv:1206.6896 [hep-th].


\bibitem{Giombi}
  S.~Giombi and V.~Pestun,
  "The 1/2 BPS 't Hooft loops in N=4 SYM as instantons in 2d Yang-Mills",
  arXiv:0909.4272 [hep-th].

\bibitem{Drukker}
  N.~Drukker,
  "The 1/4 BPS circular loops, unstable world-sheet instantons and the matrix model",
  arXiv:0605151 [hep-th].

\bibitem{Cardinali:2012sy}
  V.~Cardinali, L.~Griguolo and D.~Seminara,
  "Impure Aspects of Supersymmetric Wilson Loops,"
  arXiv:1202.6393 [hep-th].

\bibitem{Berkovi2}
N.~Berkovits,"Covariant quantization of the superparticle using pure spinors", JHEP {\bf 09} (2001) 016, [arXiv:~hep-th/0105050].

\bibitem{Olive:1995sw}
  D.~I.~Olive,
  "Exact electromagnetic duality,"
  Nucl.\ Phys.\ Proc.\ Suppl.\  {\bf 45A} (1996) 88
    [arXiv:hep-th/9508089].

\bibitem{Drukker:1999zq}
  N.~Drukker, D.~J.~Gross and H.~Ooguri,
  "Wilson loops and minimal surfaces,"
  Phys.\ Rev.\  D {\bf 60}, 125006 (1999)
  [arXiv:hep-th/9904191].


\bibitem{Migdal:1975zg}
  A.~A.~Migdal,
  "Recursion Equations in Gauge Theories,"
  Sov.\ Phys.\ JETP {\bf 42} (1975) 413
   [Zh.\ Eksp.\ Teor.\ Fiz.\  {\bf 69} (1975) 810].

\bibitem{Kazakov:1980zi}
  V.~A.~Kazakov and I.~K.~Kostov,
  "Nonlinear Strings In Two-dimensional U(infinity) Gauge Theory,"
  Nucl.\ Phys.\ B {\bf 176} (1980) 199.

\bibitem{Rusakov:1990rs}
  B.~E.~Rusakov,
  "Loop averages and partition functions in U(N) gauge theory on two-dimensional manifolds,"
  Mod.\ Phys.\ Lett.\ A {\bf 5} (1990) 693.

\bibitem{Bassetto:1998sr}
  A.~Bassetto and L.~Griguolo,
  "Two-dimensional QCD, instanton contributions and the perturbative Wu-Mandelstam-Leibbrandt prescription,"
  Phys.\ Lett.\ B {\bf 443} (1998) 325
  [hep-th/9806037].


\bibitem{poss}
  L.~F.~Alday, D.~Gaiotto, S.~Gukov, Y.~Tachikawa and H.~Verlinde,
  JHEP {\bf 1001} (2010) 113
  [arXiv:0909.0945 [hep-th]].

\bibitem{Drukker:2009id}
  N.~Drukker, J.~Gomis, T.~Okuda and J.~Teschner,
  JHEP {\bf 1002} (2010) 057
  [arXiv:0909.1105 [hep-th]].

\bibitem{Drukker:2010jp}
  N.~Drukker, D.~Gaiotto and J.~Gomis,
  JHEP {\bf 1106} (2011) 025
  [arXiv:1003.1112 [hep-th]].

\bibitem{Borokhov:2002ib}
  V.~Borokhov, A.~Kapustin and X.~k.~Wu,
  "Topological disorder operators in three-dimensional conformal field
  theory,"
  JHEP {\bf 0211}, 049 (2002)
  [arXiv:hep-th/0206054].

\bibitem{Borokhov:2002cg}
  V.~Borokhov, A.~Kapustin and X.~k.~Wu,
  "Monopole operators and mirror symmetry in three dimensions,"
  JHEP {\bf 0212}, 044 (2002)
  [arXiv:hep-th/0207074].

\bibitem{D'Adda:1977ur}
  A.~D'Adda and P.~Di Vecchia,
  "Supersymmetry and Instantons,"
  Phys.\ Lett.\ B {\bf 73} (1978) 162.



%
%
\bibitem{mcdonald}
I.G.~MacDonald, "The Volume of a Compact Lie Group," Inventiones Mathematicae 56 (1980) 93–95.

\end{thebibliography}
\end{document}